\documentclass{jfm}

\usepackage{graphicx}
\usepackage{natbib}

\usepackage{amsmath}
\usepackage{amssymb}
\usepackage{color}

\newcommand{\be}{\begin{equation}}
\newcommand{\ee}{\end{equation}}

\newcommand{\bs}{\begin{subequations}}
\newcommand{\es}{\end{subequations}}

\newcommand{\rmd}{\mathrm{d}}
\newcommand{\rme}{\mathrm{e}}
\newcommand{\rmi}{\mathrm{i}}
\newcommand{\ba}[2]{\begin{array}{c} #1 \\ #2 \end{array}}

\newcommand{\half}{{\textstyle \frac12}}
\newcommand{\ff}{\text{f.f.}}

\newcommand{\tH}{h}

\newcommand{\tzeta}{\tilde{\zeta}}
\newcommand{\tx}{\tilde{x}}
\newcommand{\tz}{\tilde{z}}

\newcommand{\tG}{\tilde{\Gamma}}
\newcommand{\bi}[1]{{\boldsymbol #1}}

\DeclareMathOperator*{\Res}{Res}
\DeclareMathOperator{\sg}{sign}
\DeclareMathOperator{\arcoth}{arcoth}

\definecolor{red}{rgb}{0,0,0}
\definecolor{black}{rgb}{0,0,0}

\ifCUPmtlplainloaded \else
  \checkfont{eurm10}
  \iffontfound
    \IfFileExists{upmath.sty}
      {\typeout{^^JFound AMS Euler Roman fonts on the system,
                   using the 'upmath' package.^^J}%
       \usepackage{upmath}}
      {\typeout{^^JFound AMS Euler Roman fonts on the system, but you
                   dont seem to have the}%
       \typeout{'upmath' package installed. JFM.cls can take advantage
                 of these fonts,^^Jif you use 'upmath' package.^^J}%
       \providecommand\upi{\upi}%
      }
  \else
    \providecommand\upi{\upi}%
  \fi
\fi

\ifCUPmtlplainloaded \else
  \checkfont{msam10}
  \iffontfound
    \IfFileExists{amssymb.sty}
      {\typeout{^^JFound AMS Symbol fonts on the system, using the
                'amssymb' package.^^J}%
       \usepackage{amssymb}%
         \let\leq=\leqslant
       \let\ge=\geqslant  \let\geq=\geqslant
      }{}
  \fi
\fi

\ifCUPmtlplainloaded \else
  \IfFileExists{amsbsy.sty}
    {\typeout{^^JFound the 'amsbsy' package on the system, using it.^^J}%
     \usepackage{amsbsy}}
    {\providecommand\boldsymbol[1]{\mbox{\boldmath $##1$}}}
\fi

\title[Oscillating line source in shear flow]{Oscillating line source in a shear flow with a free surface: critical layer-like contributions}

\author[S. \AA. Ellingsen and P. A. Tyvand]{Simen \AA. Ellingsen$^1$\thanks{Email address for correspondence: simen.a.ellingsen@ntnu.no} and Peder A. Tyvand$^2$}
\affiliation{$^1$Department of Energy and Process Engineering, Norwegian University of Science and Technology, N-7491 Trondheim, Norway \\
$^2$Department of Mathematical Sciences and Technology, Norwegian University of Life Sciences, N-1432 \AA s, Norway}

\begin{document}

\maketitle
\begin{abstract}
The linearized water-wave radiation problem for an oscillating submerged line source in an inviscid shear flow with a free surface is investigated analytically at finite, constant depth in the presence of a shear flow varying linearly with depth. The surface velocity is taken to be zero relative to the oscillating source, so that Doppler effects are absent. The radiated wave out from the source is calculated based on Euler's equation of motion with the appropriate boundary and radiation conditions, and differs substantially from  the solution obtained by assuming potential flow. To wit, an additional wave is found in the downstream direction in addition to the previously known dispersive wave solutions; this wave is non-dispersive and we show how it is the surface manifestation of a critical layer-like flow generated by the combination of shear and mass flux at the source, passively advected with the flow. As seen from a system moving at the fluid velocity at the source's depth, streamlines form closed curves in a manner similar to Kelvin's cat's eye vortices. A resonant frequency exists at which the critical wave resonates with the downstream propagating wave, resulting in a downstream wave pattern diverging linearly in amplitude away from the source. 
\end{abstract}

\section{Introduction}

The technique of using flow singularities to model floating and submerged bodies has been enormously successful in marine hydrodynamics when irrotational (potential) flow is assumed. Since in inviscid theory the only boundary condition on a solid surface is impermeability, a streamline and a solid surface is one and the same thing. The linearity of the governing equation (Laplace equation) allows one to use a distribution of singularities to satisfy boundary conditions on a solid body, for instance using panel methods. The technique is computationally far cheaper than solving the corresponding boundary value problem with the full equations of motion while still giving satisfactory results for description of wave-body interactions \citep[see, e.g.,][]{newman77,faltinsen90}. 
The submerged oscillatory source is thus recognized as an elementary
solution for linearized water waves governed by Laplace's
equation, and the first mathematical solutions were given by 
\cite{kochin39,kochin40}; see the review article by \cite{wehausen60}.

Inspired by the success of flow singularities in irrotational flow, one might hope in time to devise a corresponding theory when a shear current is present. Now the equations of motion do not reduce to the Laplace and Bernoulli equations, yet provided the inserted flow singularities are presumed weak, a theory to linear order in perturbations still permits superposition, giving hope that submerged oscillating sources could provide the desired building bricks from which a theory could be constructed for free-surface motion of bodies in shear flow. 

As a first step towards such a theory, it is necessary to fully understand the fundamental properties of the building bricks themselves. At present, however, very little is known about the behaviour of oscillating sources in the presence of shear even in two dimensions, and in three dimensions no previous work exists at all to our knowledge. 

Recently, \cite{tyvand14} found solutions for the 2D radiation problem for a submerged oscillatory line source in shear flow with a free surface without zero surface velocity, working under the assumption that the problem could be described in the framework of the Laplace equation. The solution was soon extended
to nonzero surface velocity relative to the source \citep{tyvand15}, so that Doppler effects must also be accounted for, increasing the number of far-field waves and making resonance possible.

The assumption made in these works that the perturbative motion obeys the Laplace equation is mathematically convenient and rests on the assumption that although vorticity is present in the background flow, the additional wave motion must be irrotational. This may be argued for by noting that
vorticity is conserved in two dimensions because of Lord Kelvin's circulation theorem. However, this is an uncertain argument in the the presence of a submerged source that violates Laplace's equation in one singular point. 
We have therefore revisited the oscillating source problem using a fundamental approach based on Euler's equations of motion instead of Laplace's equation, and the results obtained do not agree with the analysis of \cite{tyvand14}. 
Indeed, a simple argument can be made based on the vorticity equation which shows that this must be so, which we give in Sec.~\ref{sec:introvort}. Our present flow model based on the Euler equation of motion is clearly superior to the potential flow model of \cite{tyvand14} because it follows from first principles. Nevertheless the predictions based on potential theory give a useful background for discussion.

We solve the radiation problem for an oscillating line source in a shear flow which is at rest relative to the undisturbed surface, so that Doppler effects are absent. It is demonstrated
how the potential theory for water waves with uniform vorticity has 
shortcomings even for strictly 2D flows if there are singularities in the fluid domain. For 3D water waves, it is commonly known that they must carry a varying perturbation vorticity in the presence of a shear flow \citep{constantin11b, ellingsen16}. 
In a paper submitted alongside the present one, we have solved the sibling problem of an oscillating point source in a fully 3D flow \cite{ellingsen15b}. 

When subjecting the problem to the full treatment using the Euler equations, linearized with respect to perturbative quantities, the resulting wave pattern and velocity field can differ greatly from that obtained with potential theory. 
To wit, the presence of an oscillating source creates a flow pattern similar to a critical layer of vortical cat's eye-like structures, giving rise to a wave-like surface elevation which is not dispersive but is convected downstream at the velocity of the fluid at the source position. 

For convenience we shall often refer to the string of vortices downstream of the oscillating source simply as its ``critical layer", although the reader should note an important difference between the present flow structure and critical layers as they most often appear in, e.g., atmospheric sciences (see brief literature review below): the generated vortices drift downstream at the local flow velocity which is \emph{not} in general equal to the phase velocity of propagating waves. This is because the critical layer is generated by the source at the source's level, not by a propagating surface wave impinging on a critical level where its phase velocity equals the flow velocity. Rather than being created by an incoming propagating wave, ``our'' critical layer creates its own surface elevation which we call the critical wave, which is in general independent of the propagating wave. In the special case where the critical layer velocity \emph{does} equal the phase velocity at the forcing frequency, a resonance occurs, and wave amplitudes grow without bound (in the linearised theory). We return to this point in Sec.~\ref{sec:critlayer}. 

It has been shown recently in the mathematical literature \citep{ehrnstrom08,wahlen09} that the linearized 2D Euler equation 
with constant vorticity
can support critical layer solutions in which streamlines can form closed loops (or ``vortices'') as seen from a system moving downstream with the same velocity as the critical layer is convected,
confirming a classical conjecture by Lord Kelvin \citep{kelvin1880}. Mathematical studies of critical layers for internal gravity waves have since become even more wide-reaching, and a review of recent progress, including the use of elegant conformal mapping techniques, is given by \cite{constantin11}.

Critical layers have been much discussed in connection with the generation of surface waves by wind. The possibility was first investigated by \cite{miles57}, who pointed out that a critical layer in the air layer above the water surface is able to impart horizontal momentum to the surface waves, thus exciting gravity waves at much lower air velocities than predicted by the classical Kelvin-Helmholtz instability theory. The problem was recently revisited by \cite{walsh13,buhler15}. For internal gravity waves in water, critical layers are known to be able to strongly attenuate the amplitude of internal gravity waves as they pass through the layer \cite{booker67, thorpe81}. Herein, only the generation of a critical layer is studied; an interesting and potentially important problem not pursued here concerns the stability and eventual fate of the critical layer. In particular, upon re-introducing a small viscosity near the critical layer, Tollmien--Schlichting waves are known to occur, providing a potential mechanism for generation of turbulence \citep[e.g.,][]{maslowe86}. Indeed, the eventual fate of a perturbed vortex is a question much studied in recent times, e.g., by \cite{balmforth01}, which lies beyond the scope of our present analysis.

A rich literature exists regarding waves with uniform vorticity in 2D. The classical theory of waves on shear currents and critical layers is well summarised in monographs by \cite{drazin81}, \cite{craik85} and \cite{buhler09}, as well as the classic reviews by \cite{peregrine76} and \cite{peregrine83}.  Nonlinear waves were considered by \citet{telesdasilva88} and \cite{kang00}, and recently the Benjamin-Feir instability was investigated for weakly non-linear waves on a shear current by \cite{thomas12} with applications to the field of giant waves. 
A large body of recent work on large-amplitude waves was spurred by the progress of \cite{constantin04}, including the striking result that steady, periodic waves on a shear flow must be symmetrical about its crest regardless of the flow vorticity \citep{constantin04b,constantin07, tulzer12}, with accompanying numerical demonstration \citep{ko08,constantin15}.

\section{Physical system}

We here describe the physical system under consideration, and discuss its properties with respect to vorticity, which immediately sets it apart from submerged sources as they appear within potential theory.

\begin{figure}
  \includegraphics[width=\textwidth]{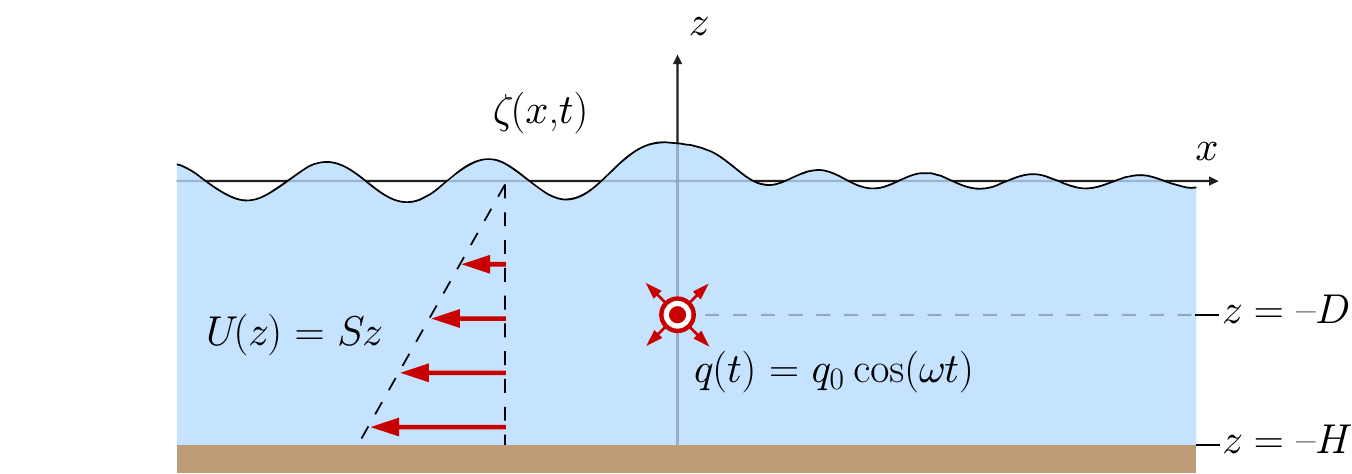}
  \caption{The geometry considered: an oscillating line source at depth $D$.}
  \label{fig:geom}
\end{figure}

We consider an inviscid and incompressible fluid in a steady shear
flow, where the shear flow is aligned along a horizontal $x$ axis.
The fluid has constant depth and a free surface subject
to constant atmospheric pressure. Cartesian coordinates $x, z$ are
introduced, where the $z$ axis is directed upwards in the gravitational
field, and the $x$ axis is aligned along the undisturbed free
surface. The gravitational acceleration is $g$, and $\rho$ denotes
the constant fluid density. 
Surface tension is neglected.
The velocity perturbation vector is
denoted by 
$ \hat{\bi{v}}=(\hat{u},\hat{w})$. 
The surface elevation is
denoted by $\zeta(x,t)$, and the overall problem is sketched in
Figure \ref{fig:geom}. There is a wave motion driven by a fixed oscillating line source located in the point $(x,z)=(0,-D)$. The water wave problem will be linearized with respect to the surface elevation
and the velocity and pressure perturbations. 

We assume constant fluid depth $H$. There is a basic horizontal shear flow $U(z)$ in the $x$ direction
\begin{equation}\label{eq:shearVel}
U(z) = U_0 + S z,~~z <0,
\end{equation}
where $U_0$ is the uniform surface velocity, which will later be put equal to zero. There is a uniform vorticity $S$ in the 
$y$ direction. 
We make the restriction $S \ge 0$.

We assume time-periodic flow with given, constant angular frequency $\omega>0$. This
flow is driven by a point source of harmonically pulsating strength
\begin{equation}\label{eq:source}
  q(t) = 
  q_0 e^{-\rmi\omega t}.
\end{equation}
The source strength $q(t)$ is the instantaneous volume flux per unit length
emitted from the singular source, which is located at the constant depth $z=-D$. Here and elsewhere the physical quantity is the real part.
The continuity equation is thus inhomogeneous and reads:
\be\label{eq:sourcepoint}
  \nabla\bi{\cdot} \hat{\bi{v}} = q_0 \delta(x) \delta(z+D) \rme^{-\rmi \omega t}.
\ee

\subsection{Vorticity generation}\label{sec:introvort}

Before going on to solve the wave problem just laid out, let us briefly discuss a key property which sets the oscillating source submerged in a shear current apart from its better known sibling problem in irrotational flow, namely the conservation of vorticity.

\cite{tyvand14} considered the problem assuming potential theory, based on the notion that the vorticity of each fluid particle is conserved in inviscid 2D flow. This is not necessarily the case in the presence of flow singularities, however. The vorticity equation for 2D flow, governing the time development of vorticity $\Omega$ (directed in the $y$ direction) reads
\be\label{eq:vort}
  \frac{\mathrm{D}\Omega}{\mathrm{D}t} = -\Omega (\nabla \bi{\cdot} \hat{\bi{v}}) = -S q_0\delta(x) \delta(z+D) \rme^{-\rmi \omega t}
\ee
where we insert \eqref{eq:sourcepoint} and keep terms to linear order.

From \eqref{eq:vort} it is immediately clear that the line source will change the vorticity of the flow. A fluid particle passing through the point $(0,-D)$ is given an additional vorticity due to the term on the right hand side of \eqref{eq:vort}. The time the particle spends in the vicinity of the source is proportional to $1/U(-D)$, hence the additional vorticity picked up by such a particle and carried downstream is expected to be proportional to $-Sq_0\exp(-\rmi\omega t)/U(-D)$, and we must expect a thin downstream line of vorticity to appear, oscillating in sign with a wavelength of $2\upi |U(-D)|/\omega$, the distance travelled by such a particle during one oscillation period. Indeed, we will show this to be the case in later sections. 

It is clear that this injection of vorticity into the flow might complicate the use of oscillating sources as Green functions for the linearised free surface Euler equations, as a means to model the behaviour of a submerged body. The downstream vorticity sheet would make boundary conditions difficult to satisfy and would moreover be an unacceptable model of a real body, which can never inject additional vorticity into an inviscid flow due to Kelvin's circulation theorem. 
The question is discussed further in Section \ref{sec:critfree} where a possible remedy is indicated.

Before any such application can be undertaken, however, it is necessary to study the oscillating source model and lay out its properties such as they are. Equation \eqref{eq:vort} shows that the injection of vorticity is a necessary property of any submerged source in a shear flow, and the corresponding additional solution to the wave problem found in the following is as such not a spurious solution, but one that must be accounted for and treated consistently. 

\subsection{Equation of motion and boundary conditions}

We continue now to solving the wave problem.
Euler's equation of motion may be written
\begin{equation}\label{eq:euler}
\bi{a} = - \frac{1}{\rho} \nabla P - g \bi{e}_z,
\end{equation}
where $\bi{a}$ is the acceleration vector and $\bi{e}_z$ is the vertical unit vector. The total pressure is denoted by $P = -\rho g z + \hat p$ with $\hat p$ the small perturbation.

The linearized kinematic free-surface condition is
\begin{equation}\label{eq:surfkincond}
\left. \hat{w} \right\vert_{z=0} = \zeta_t + U_0 \zeta_x,
\end{equation}
with subscripts denoting partial derivatives.

The dynamic boundary condition is given by the continuity of the tangential component of the Euler equation along the free surface. It can be written
\begin{equation}\label{eq:surfcond}
\bi{a} - (\bi{a}\cdot\bi{n})\bi{n} = - g \bi{e}_z + g (\bi{e}_z\cdot\bi{n})\bi{n},~~z=\zeta(x,y,t),
\end{equation}
where the surface normal vector is denoted by $\bi{n}$. According to linear theory, the surface normal is given by $\bi{n}=\bi{e}_z-\nabla \zeta$. We linearize this dynamic free-surface condition and take its $x$ component, which gives
\begin{equation}\label{eq:surfcond1}
\hat{u}_t + U_0 \hat{u}_x + S \hat{w} = - g \zeta_x,~~z=0.
\end{equation}
From now on, we will put $U_0 = 0$ in the present analysis. This entails that the source is now stationary as seen by the moving surface, eschewing additional complications from Doppler effects \citep[cf., e.g.,][]{tyvand15}. Only the relative velocity between source and free surface is of importance, and an overall constant velocity can be transformed away by a change of coordinate system. 
Since the flow is incompressible we eliminate the horizontal velocity to get the dynamic free-surface condition
\begin{equation}\label{eq:dynamic}
\hat{w}_{zt} - S \hat{w}_x =  g \zeta_{xx},~~z=0,
\end{equation}
expressed in terms of the vertical velocity and the elevation.

The last boundary condition is the bottom condition
\begin{equation}\label{eq:bottom}
\hat{w} = 0,~~z=-H.
\end{equation}

\subsection{Fourier transform of the radiation problem}

The variables are Fourier transformed as follows
\begin{equation}\label{eq:fourier}
  (\hat{u},\hat{w},\hat{p}) =  q_0 \frac{1}{2 \upi} \int_{-\infty}^{\infty}\rmd k\,
  (u(z),w(z),p(z)) e^{\rmi  k  x - \rmi  \omega t}.
\end{equation}
This Fourier integral consists of two different contributions: Waves propagating in the $+x$ direction, with positive wavenumber ($k>0$), and waves propagating in the $-x$ direction, with negative wavenumber ($k<0$). This is a contrast to the potential theory for the same problem  by
\citet{tyvand14}, where it was convenient to work only with positive wave numbers.

The transformed components of the Euler equation are given by
\bs
\begin{align}\label{eq:euler1}
  - \rmi  (\omega - k U) u + S w =& - \rmi  k p/\rho, \\
  \label{eq:euler3}
  - \rmi  (\omega - k U) w  =& - p'/\rho,
\end{align}
\es
where a prime denotes a derivative taken with respect to $z$. The transformed continuity equation \eqref{eq:sourcepoint} is
\begin{equation}\label{eq:continuity}
  \rmi  k u + w' = \delta (z+D).
\end{equation}
The surface elevation is Fourier transformed as follows
\begin{equation}\label{eq:elevfourier}
  \zeta(x,t) = q_0 \frac{1}{2 \upi} \int_{-\infty}^{\infty}\rmd k\, B(\omega,k)  e^{\rmi  k x  - \rmi  \omega t}.
\end{equation}

\section{Solution for the submerged line source}\label{sec:solution}

We will now solve the radiation problem for the line source for finite depth. From the
transformed governing equations we derive an inhomogeneous Rayleigh equation for the
vertical velocity alone
\begin{equation}\label{eq:vertfourier}
w'' - k^2 w = \delta'(z+D)-\frac{k S}{\omega - k U}\delta(z+D).
\end{equation}
Since the $\delta$ function is nonzero only in the point $z=-D$, the fraction $kS/(\omega-kU)$ effectively becomes $kS/(\omega + kSD)$ in the last term.
The homogeneous solution of this equation is written in the form
\begin{equation}\label{eq:homo}
w_h(z) =  A(\omega,k) \sinh k(z+H),
\end{equation}
satisfying the bottom condition (\ref{eq:bottom}).
In addition, there are two inhomogeneous solutions, due to the two forcing terms in (\ref{eq:vertfourier})
\begin{align}\label{eq:inhomo1}
w_{p1}(z) =&  \cosh k(z + D) \Theta(z + D),\\
\label{eq:inhomo2}
w_{p2}(z) =& -\frac{S}{\omega + k S D} \sinh k(z +
D) \Theta(z + D),
\end{align}
where the Heaviside unit step function $\Theta(z)$ has been introduced. 

A notable advantage achieved by working with finite depth $H$, is that we can develop the solution without paying attention to the sign of $k$. With infinite depth, the homogeneous solution of (\ref{eq:vertfourier}) would have to be written $w_h(z) = A(\omega,k) \exp (\vert k \vert z)$.

From transforming the kinematic free-surface condition (\ref{eq:surfkincond}) we find the relationship
\begin{equation}\label{eq:boga}
A \sinh kH  = - \cosh (kD) + \frac{S \sinh (kD)}{\omega + k S D} -\rmi  \omega B.
\end{equation}
Similarly, the dynamic free-surface condition (\ref{eq:dynamic}) gives
\begin{align}\label{eq:aogb}
(\omega  \coth kH  +& S ) A \sinh kH + \omega \sinh (kD) + S \cosh (k D)\notag \\
 -&
\frac{\omega S \cosh (kD) +  S^2 \sinh (kD)}{\omega + k S D} = - \rmi  g k B.
\end{align}
From the two last equations we eliminate $A$, since we are most interested in the surface elevation.
It is expressed by $B(\omega,k)$, given as
\begin{align}\label{eq:bformel0}
\rmi  \frac{g}{\omega} \left(k - \frac{\omega (\omega \coth kH + S)}{g}\right) B
=& \frac{\cosh k(H-D)}{\sinh kH} 
+ \frac{\sinh k(H-D)}{(k D + \omega/S)\sinh kH} .
\end{align}
Here $B$ is multiplied by a factor that will be zero when $k$ has
one of the two values corresponding to the
dispersion relation for given $\omega$. The far-field wave
propagating in the $+x$ direction has positive wavenumber $(k=k_+>0)$, while
the far-field wave propagating in the $-x$ direction has negative
wavenumber $(k=k_-<0$). The dispersion relation has an implicit form
\begin{equation}\label{eq:disprelasjon}
  g k = \omega (\omega \coth kH + S),
\end{equation}
from our point of view, where we want to determine the values $k=k_\pm$ satisfying $k=k(\omega)$   for a given value of $\omega$. For all angular frequencies $\omega>0$ there is one positive solution $k_+$ and one negative solution $k_-$. Their absolute values are given by the implicit relationships
\begin{equation}\label{eq:disprell}
  g k_+ = \omega (\omega \coth k_+ H + S),~~~~g \vert k_- \vert = \omega (\omega \coth \vert k_- \vert H - S),
\end{equation}
where there is one positive solution for $k_+$ and one positive solution for $\vert k_-\vert$. These two formulas share
the classical limit $g k_\pm \tanh k_\pm H = \omega^2$ as the shear flow vanishes $(S \rightarrow 0)$.

The dispersion relation (\ref{eq:disprelasjon}) has been derived by 
\citet{tyvand14}, with a different sign convention for $k_-$. They emphasized the cut-off that exists for the downstream wave ($k$ negative by the convention) when $\omega<S$ in the limit $H \rightarrow \infty$. However, this cut-off is an artifact that does not exist for finite $H$, no matter the value of $H$. We discuss this point also in the below when regarding the deep water limit of our results. 
See also the more detailed discussion of the dispersion relation by 
\cite{ellingsen14c}, 
who took the conventional point of view that
the wave frequency varies with the wave number.

A brief comment is warranted at this stage concerning the apparently odd fact that the Fourier transformed vertical velocity $w$ is discontinuous at $z=-1$. Clearly the physical quantity $\hat{w}$ must be continuous, and indeed it is, everywhere except in the single point $(0,-D)$, the position of the source. Since the Fourier transform of $\delta(x)$ is $1$, the discontinuity at $x=0$ is present for all $k$ at the level $z=-1$. This is no problem in principle, but requires some care when numerical calculations are performed in Sec.~\ref{sec:velocity}.

\section{Radiation condition and far-field solution}
\label{sec:ff}

The wave pattern as given by \eqref{eq:elevfourier} with $B$ from \eqref{eq:bformel0} is not yet well defined because it has poles sitting exacly on the axis of integration. This complication is well known and results from a purely periodic model not having an arrow of time, hence being unable to distinguish between past and future. To remedy this in a manner which is satisfactory both mathematically and physically, we enforce a radiation condition using the method advocated by, \emph{inter alia}, Lighthill \citep[e.g.,][\S 4.9]{lighthill78}, whereby we suppose that the oscillating source has been increasing slowly in strength since $t=-\infty$, so that \eqref{eq:sourcepoint} is replaced by
\be
  q(t) = q_0e^{-\rmi \omega t + \epsilon \omega t}
\ee
where the limit $\epsilon\to 0^+$ will eventually be taken.
This amounts to the replacement everywhere
\be
  \omega \to \omega(1 + \rmi \epsilon ).
\ee
In the limit $\epsilon \to 0^+$, the only remaining effect of the replacement is that the poles on the $k$ axis are now moved slightly into the complex $k$ plane, rendering the integral well defined. Re-writing \eqref{eq:bformel0} slightly and expanding to leading order in $\epsilon$ near the relevant zeros of denominators, we can express the surface elevation as
\begin{align}\label{eq:zfull}
 \frac{g}{q_0\omega} \zeta(x,t) =& -\rmi \lim_{\epsilon\to 0^+}\int_{-\infty}^\infty\frac{\rmd k}{2\upi} \frac{e^{\rmi kx-\rmi \omega t}}{[\Gamma(k)-\rmi \epsilon\Phi(k)]\sinh kH}\notag \\
 &\times\left[\cosh k(H-D) + \frac{\sinh k(H-D)}{D(k-k_c-\rmi \epsilon k_c)} \right]
\end{align}
where we have defined
\bs
\begin{align}
  \Gamma(k) =& k -\frac{\omega^2}{g}\coth kH - \frac{\omega S}{g} , \\
  \Phi(k) =& \frac{2\omega^2}{g}\coth kH + \frac{\omega S}{g},\\
  k_c =& -\frac{\omega}{SD}.
\end{align}
\es
Equation \eqref{eq:zfull} is the complete expression for $\zeta$ and may be evaluated numerically as it is, using a small but nonzero value of $\epsilon$ which amounts to a slight attenuation of the waves away from the source. This will give the full surface elevation including both near field and far-field solution, and is thus a useful check for our asymptotic results in the following. 

In order to evaluate integral \eqref{eq:zfull} using residue calculation we must find out where the poles of the integrand lies. For shorthand, let us first write \eqref{eq:zfull} as
\[
  \frac{g}{q_0\omega} \zeta(x,t) = \lim_{\epsilon\to 0^+}\int_{-\infty}^\infty \rmd k f(k).
\]
Inspecting $f(k)$ for complex $k$ we see that it has poles of three distinct types. Firstly, there are the poles near $k=k_\pm$ where $\Gamma(k)=0$. These are the values of $k$ which satisfy the the dispersion relation, as discussed above. Next there is a pole at $k=k_c$ which we will discuss further below. It stems from the formation of a 
critical layer-like vortical flow structure downstream of the source, 
a phenomenon which cannot be described by potential theory, and is therefore not present in the solutions of \citet{tyvand14}. Finally one finds that $\Gamma(k)$ has an infinite number of zeros lying somewhat to the right of the imaginary axis (they move onto the imaginary axis as $S\to 0$), none of which are close to the real axis. These provide near field contributions to the integral. 

The poles near $k=k_\pm$ and $k=k_c$ have been moved slightly off the real axis by the introduction of $\epsilon$. Their new positions are found at
\be
  k = k_\pm + \rmi \epsilon \frac{\Phi(k_\pm)}{\Gamma'(k_\pm)} + \mathcal{O}(\epsilon^2)
\ee
where $\Gamma' = \rmd \Gamma/\rmd k$ and only the leading order in $\epsilon$ is kept.
We have
\be
  \Gamma'(k_\pm) = 1+ \frac{\omega^2H}{g \sinh^2 k_\pm H} 
\ee
which is clearly a positive function for all $k_\pm$. Hence the pole at $k_\pm$ is moved slightly above the real axis if $\Phi(k_\pm)>0$ and below if $\Phi(k_\pm)<0$. Inserting $
  \frac{\omega^2}{g}\coth k_\pm H = k_\pm - \frac{\omega S}{g}$
from equation 
(\ref{eq:disprell}) 
we can write
\be
  \Phi(k_\pm) = 2k_\pm - \frac{\omega S}{g}.
\ee
Since $k_-<0$ it is clear that $\Phi(k_-)<0$. Apparently $\Phi(k_+)$ can take both signs, but this would require that $k_+ < \omega S/2g$, and one can show that no such solution to the dispersion relation exists. Hence the pole at $k_+$ is moved above the $k$ axis and that at $k_-$ is moved below. Obviously the pole at $k=k_c(1+\rmi \epsilon)$ now sits below the axis, since $k_c<0$. 

\begin{figure}
    \includegraphics[width=\textwidth]{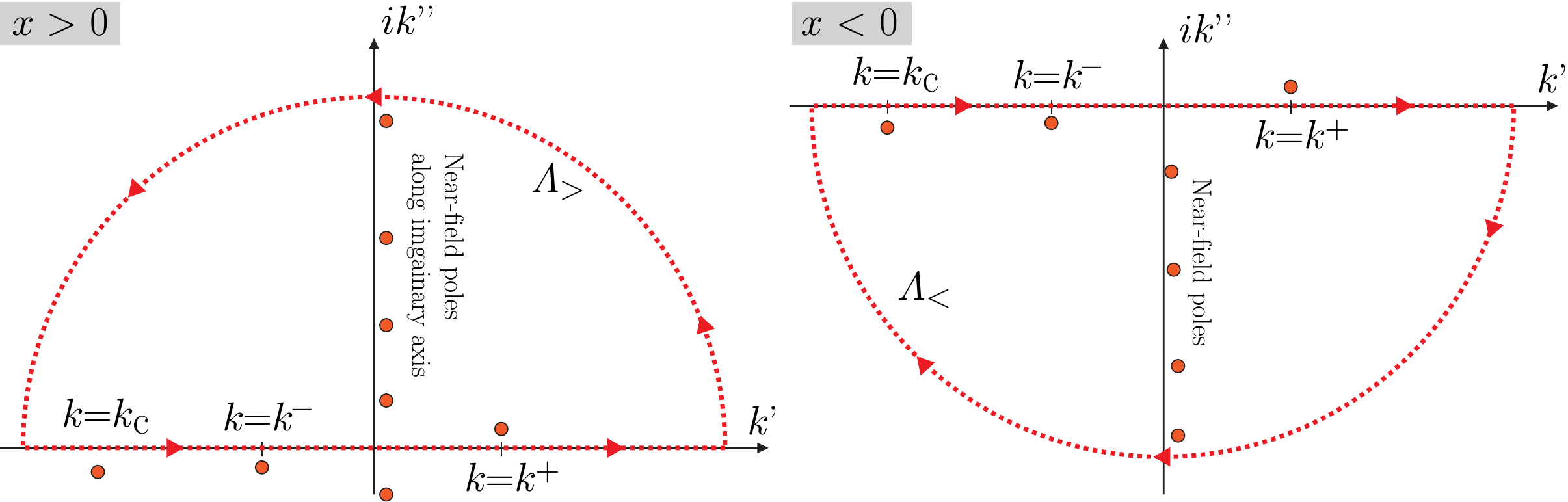}
    \caption{Integration contours and poles in the complex $k$ plane, where $k=k'+ik''$.}
    \label{fig:contours}
\end{figure}

When $x>0$ we must close the integration contour in the upper complex plane in order for the closing path (a large semicircle at infinity) to vanish. The opposite is the case for $x< 0$. 
On the semicircular arc at $|k|\to \infty$, the integrand 
$f(k)$ is exponentially small, 
hence
we get far-field contributions to $\oint \rmd k f(k)$ from the pole near $k_+$ for $x>0$, and from $k_-$ and $k_c$ downstream, at $x<0$. Poles and contours are sketched in figure \ref{fig:contours}. 

Thus we can evaluate the integral for the surface elevation for $x>0$
\begin{align}
  \frac{g}{q_0\omega} &\zeta(x,t) = \oint_{\Lambda_>} \rmd k f(k) = 2\upi \rmi  \Res_{k=k_+}f(k) + \text{near-field}\notag \\
  =& \frac{e^{\rmi k_+x-\rmi \omega t}}{\Gamma'(k_+)\sinh k_+H} \left[\cosh k_+(H-D) + \frac{\sinh k_+(H-D)}{D(k_+-k_c)} \right] 
  + \text{near-field}
\end{align}
and for $x<0$ (where we get the opposite sign since poles are encircled in the negative sense)
\begin{align}
  \frac{g}{q_0\omega}& \zeta(x,t) = \oint_{\Lambda_<} \rmd k f(k) 
  = -2\upi \rmi  \Res_{k=k_-}f(k)-2\upi \rmi  \Res_{k=k_c}f(k) + \text{near-field}\notag \\
  =& -\frac{e^{\rmi k_-x-\rmi \omega t}}{\Gamma'(k_-)\sinh k_-H} \left[\cosh k_-(H-D) + \frac{\sinh k_-(H-D)}{D(k_--k_c)} \right] \notag\\
  &- \frac{e^{\rmi k_c x - \rmi \omega t}}{D\Gamma(k_c)}\frac{\sinh k_c(H-D)}{\sinh k_c H} + \text{near-field}.\label{eq:zfdownstream}
\end{align}
The last of these explicit terms is the far-field elevation of the critical wave. We note that the factor $\Gamma(k_c)$  depends on gravity, while no other factors affecting the critical wave involve gravity. The critical wave has a weaker link to gravity than the two regular wave components; it is the surface manifestation of what is primarily a subsurface phenomenon, modified by but not caused by gravity.

For the purposes of evaluating the far-field amplitudes we use the following expression
\be
  \Gamma'(k_\pm) = 1+\frac{H}{g}(S^2-\omega^2)-2\frac{S}{\omega}k_\pm H + \frac{gk_\pm^2 H}{\omega^2}.
\ee
by noting that $\sinh^{-2}x = \coth^2 x -1$ and substituting \eqref{eq:disprelasjon}.

Note, importantly, that we assume in the expression for $x<0$ that $k_- \neq k_c$. The case $k_-=k_c$ is a special case which we will consider especially below. 

Note moreover that when the shear is weak, the water can easily seem ``deep'' for the critical layer contribution even though $|k_\pm H|$ is not large. This is the case when $|k_cH|\gtrsim 2$, or
\be\label{eq:weakshear}
  H_S=\frac{\omega H}{SD}\gtrsim 2.
\ee
The quantity $H_S$ could be called the dimensionless shear depth parameter.
The asymptote of large $H_S$ is not to be considered a deep water limit in the normal sense, and will most often come into play when $S$ is significantly smaller than $\omega$ even though the dispersive waves $k_\pm$ see shallow water dispersion.
In this case the last term of the downstream far-field \eqref{eq:zfdownstream}, the critical layer contribution, quickly approaches the simpler limiting form
\[
  \frac{e^{k_cD}e^{\rmi k_c x - \rmi \omega t}}{\omega/S -\omega^2D/g + \omega SD/g}.
\]

\subsection{Special case: $k_-=k_c$}

We consider especially the far field for $x<0$ in the special case where $k_-=k_c$. In this case the poles from the dispersion relation and the critical layer flow together, and the integrand has a double pole. The two waves coincide when
\be\label{eq:special}
  F_S^2 = \frac{1}{\coth H_S - \sigma}
\ee
where $F_S =S\sqrt{D/g}$ is a Froude number based on depth $D$ and the current velocity $S D$ at depth $D$, and $\sigma = S/\omega$. When the shear depth $H_S\gtrsim 2$, $F_S^2=1/(1-\sigma)$ is the approximate criterion.

To calculate the combined wave in the far field, we need the residue
\begin{align}
  2\upi \rmi\Res_{k=k_c}f(k)=&\Res_{k=k_c}\frac{\gamma(k)}{\Gamma(k)(k-k_c)} = \lim_{k\to k_c}\frac{\rmd}{\rmd k}\frac{(k-k_c)\gamma(k)}{\Gamma(k)} \notag \\
  =&  \frac{\gamma'(k_c)\Gamma'(k_c)-\frac12 \Gamma''(k_c)\gamma(k_c)}{\Gamma'(k_c)^2}
\end{align}
where
\be
  \gamma(k) = \frac{(k-k_c)\cosh k(H-D)+\frac1D \sinh k(H-D)}{\sinh k H}e^{\rmi kx-\rmi \omega t}
\ee
is the nonsingular part of the integrand $f(k)$. A little algebra yields the following expressions:
\begin{align}
  \gamma(k_c) &= \frac{\sinh k_c(H-D)}{D\sinh k_c H}e^{\rmi k_cx-\rmi \omega t},\\
  \gamma'(k_c) &= \frac1D\left[H \frac{\sinh k_c D}{\sinh^2 k_cH}+\rmi x\frac{\sinh k_c(H-D) }{\sinh k_cH}\right]e^{\rmi k_cx-\rmi \omega t}
\end{align}
and
\be
  \Gamma''(k) = -\frac{2\omega^2H^2}{g}\frac{\coth kH}{\sinh^2 kH}.
\ee
The surface elevation solution for $x<0$ in this case becomes, instead of \eqref{eq:zfdownstream},
\begin{align}
  \frac{g}{q_0\omega}& \zeta(x,t) = \oint_{\Lambda_<} \rmd k f(k) = -2\upi \rmi  \Res_{k=k_c}f(k) + \text{near-field}\notag \\
  =& -\frac{\gamma'(k_c)\Gamma'(k_c)-\frac12 \Gamma''(k_c)\gamma(k_c)}{\Gamma'(k_c)^2}+ \text{near-field}.\label{eq:zfdowndbl}
\end{align}
This complicated expression simplifies significantly in the weak shear asymptote \eqref{eq:weakshear}, in which case it becomes
\be
  \frac{g}{q_0\omega} \zeta(x,t) \sim -\rmi \frac xDe^{k_cD}e^{\rmi k_cx-\rmi \omega t}.\label{eq:zfweakshear}
\ee
In other words the wave amplitude grows linearly away from the source in the downstream direction in the special case where the two downstream wave numbers equal each other. One may verify that the double pole result equals the limiting value of the downstream \eqref{eq:zfdownstream} wave when $\kappa_c\to\kappa_-$.

\subsection{Limiting expressions for deep water}

Let us assume $|kH|\gg 1$ under the integral sign of Eq.~\eqref{eq:zfull}, which is unproblematic since the integrand is analytical near $k=0$ which is the only point at which such an assumption might be dubious as $H\to \infty$. The dispersion relation 
(\ref{eq:disprelasjon}) 
now has explicit solutions at 
\be
  k_{+}^\infty = \frac{\omega^2}{g}\Bigl( 1 + \frac{S}{\omega}\Bigr), ~~~ k_{-}^\infty = \frac{\omega^2}{g}\Bigl(\frac{S}{\omega}-1\Bigr)\Theta(\omega-S).
\ee
Here the Heaviside $\Theta$ function ensures that $k_-\leq 0$ as is always the case for finite $H$ (however large). The fact that $k_-^\infty$ has solution $0$ whenever $S>\omega$ is exactly the cutoff-phenomenon at deep water, discussed in detail by
\citet{tyvand14}. For any $S>\omega$ the dispersion relation has solutions so that $|k_-H| \sim \mathcal{O}(1)$. Inserting this into the far-field solution of \eqref{eq:zfdownstream} one finds that the dispersive term (from the pole at $k=k_-$ has amplitude proportional to $1/H$ and vanishes for deep water. This is not the case for $S<\omega$ where moderate (negative) solutions $k_-$ of the dispersion relation exists in the deep water limit producing waves of non-vanishing amplitude. 
The phenomenon of cut-off is discussed further in Appendix \ref{app:disprel}.

In the deep-water limit the far-field solution upstream ($x>0$) now reads
\be
  \frac{g}{q_0\omega}\zeta^>_\text{f.f.}(x,t) = e^{\rmi k_+x-\rmi \omega t}e^{-k_+D}\left[1+\frac1{D(k^\infty_+-k_c)}\right]
\ee
and downstream ($x<0$) assuming $k_c\neq k_-$,
\begin{align}
  \frac{g}{q_0\omega}\zeta^<_\text{f.f.}(x,t) =&e^{\rmi k_-x-\rmi \omega t}e^{k_-D}\left[1-\frac1{D(k_-^\infty-k_c)}\right]\Theta(\omega-S)
  +\frac{e^{k_cD}e^{\rmi k_c x - \rmi \omega t}}{D(k_-^\infty-k_c)}
\end{align}
where the last term is the weak shear limit above. When $k_c=k_-$, the result \eqref{eq:zfweakshear} is regained also in the deep water limit.

\section{Discussion of results}\label{sec:discussion}

\begin{figure}
  \begin{center}
    \includegraphics[width = \textwidth]{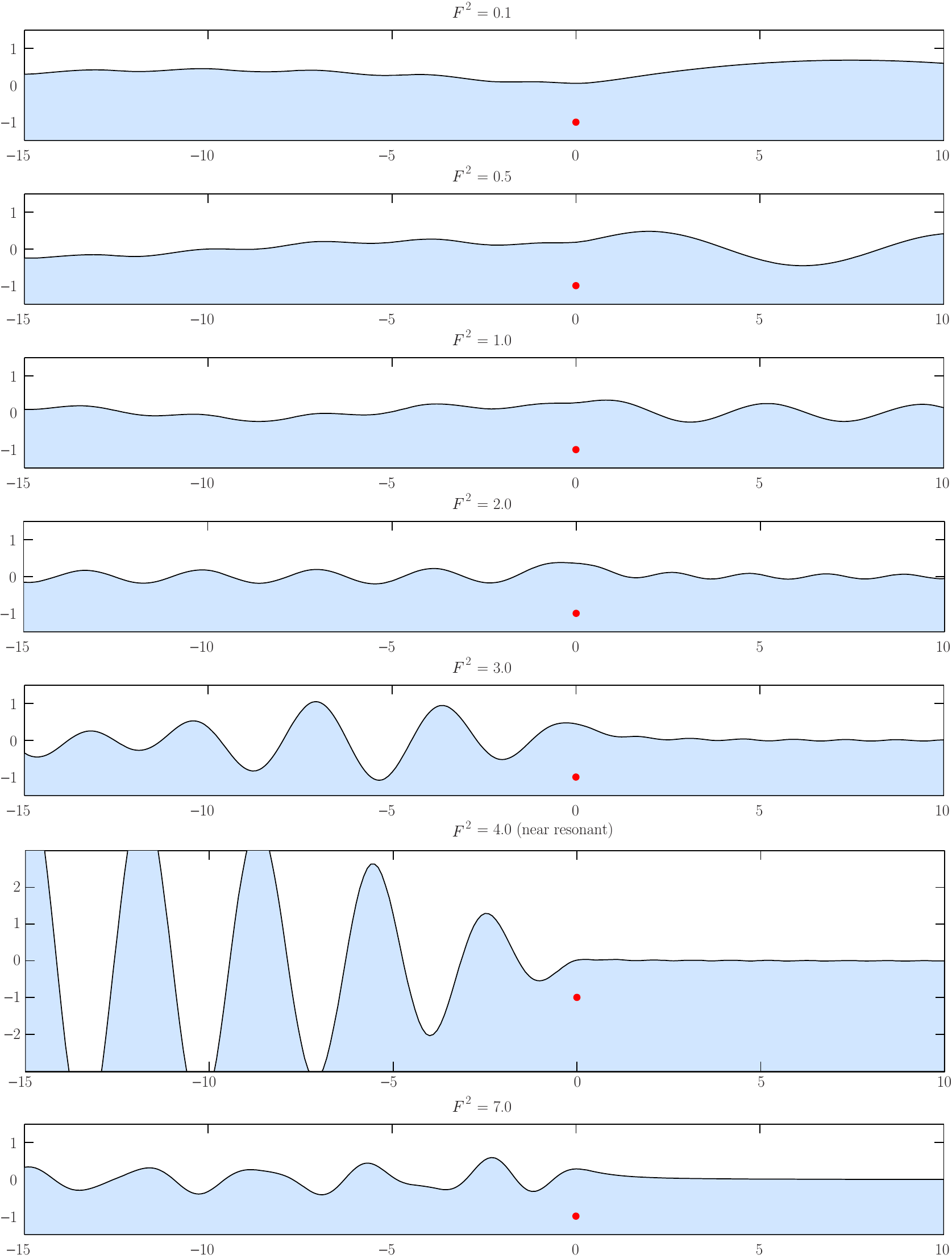}
  \end{center}
  \caption{Examples of deep water wave patterns for different values of the Froude number $F$ with parameters $\zeta_0/D=0.6$ and $\sigma=0.5$. The amplitude in each panel was multiplied by $\mathrm{min}(e^{F^2(1-\sigma)},e^{1/\sigma})$ in order to have similar downstream amplitudes. Length scales in the figure are in units of $D$. See supplementary information for video. (See also videos in supplementary materials).}
  \label{fig:panels}
\end{figure}

\begin{figure}
  \begin{center}
    \includegraphics[width = .7\textwidth]{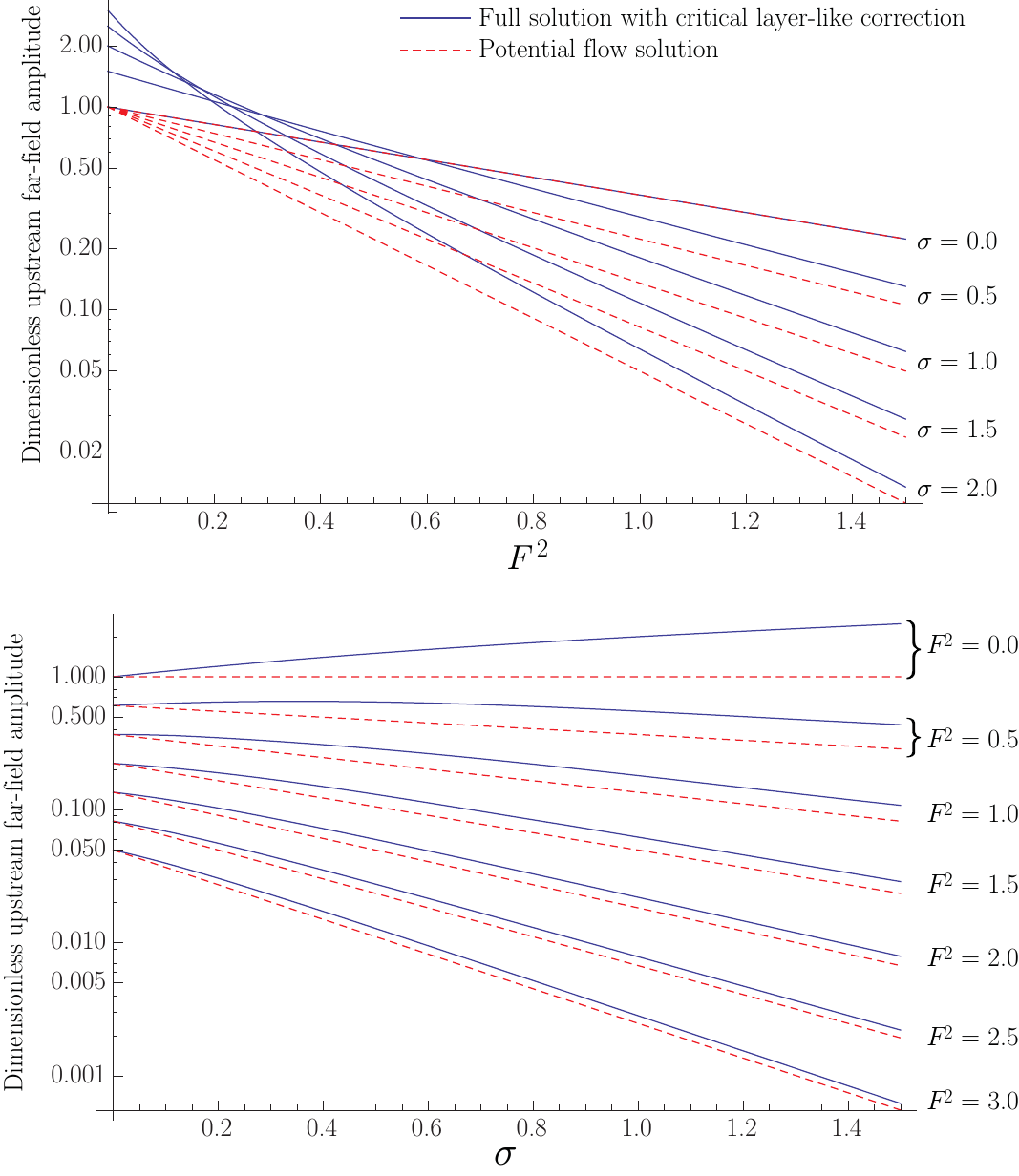}
  \end{center}
  \caption{Deep water amplitude $\tzeta$ of upstream wave as function of Froude number for different values of $\sigma$ (above) and as a function of $\sigma$ for different values of $F^2$ (below).  Potential flow results following \cite{tyvand14}.
  \label{fig:deepz}
  }
\end{figure}

\begin{figure}
  \begin{center}
    \includegraphics[width = .9\textwidth]{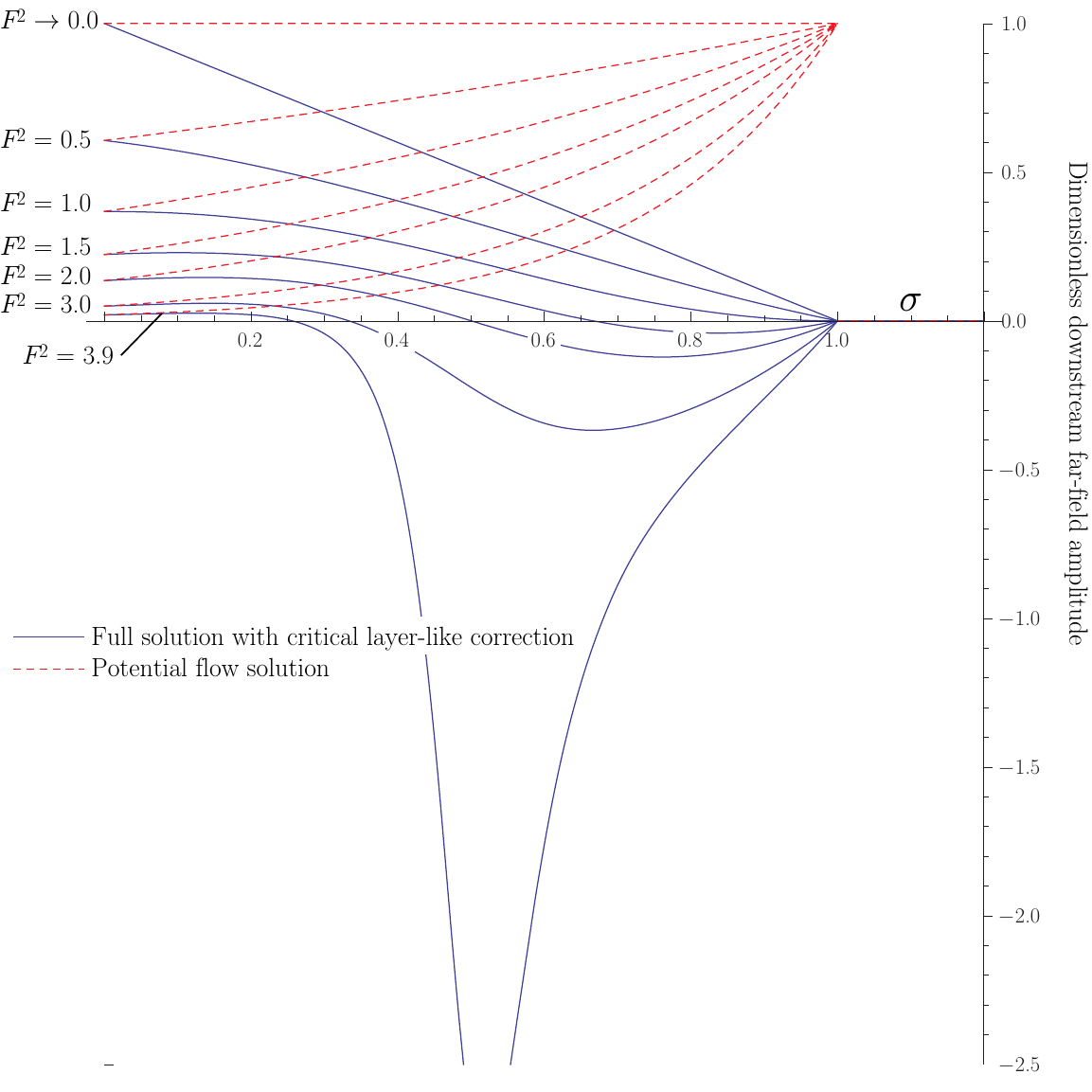} 
  \end{center}
  \caption{Deep water amplitude  $\tzeta$ of downstream regular wave as a function of $\sigma$ for different values of $F^2$. 
  Resonance can occur for $F^2\geq 4$.  Potential flow results following \cite{tyvand14}.  \label{fig:downz}
  }
\end{figure}

\begin{figure}
  \begin{center}
    \includegraphics[width = .7\textwidth]{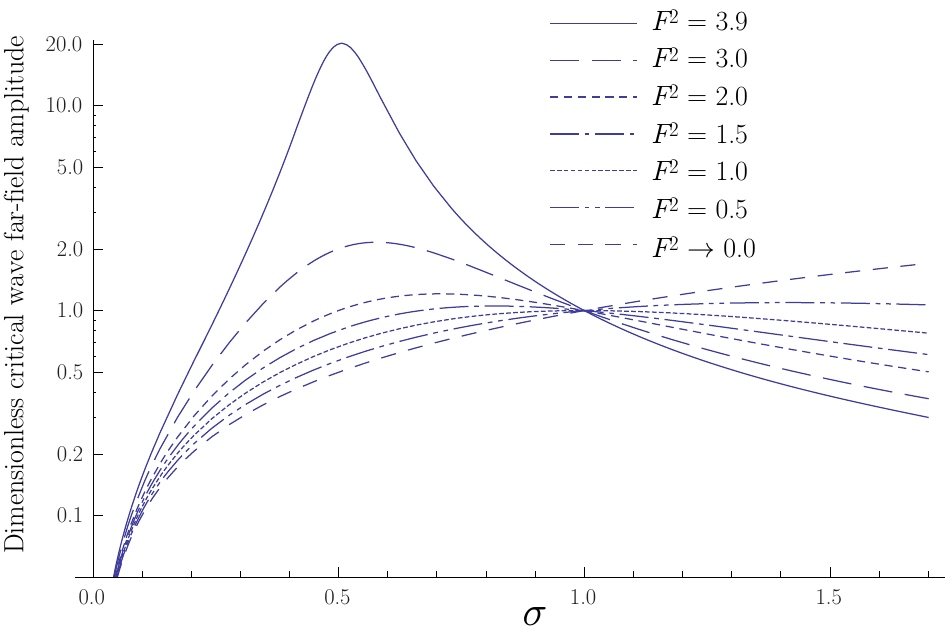} 
  \end{center}
  \caption{Deep water amplitude $\tzeta$ of critical wave as a function of $\sigma$ for different values of $F^2$.   }
  \label{fig:deepcrit}
\end{figure}

For numerical purposes we non-dimensionalise as given in Table \ref{tab:units} using $D$ and 
$\omega^{-1}$ as length and time scales, with the exception of perturbed quantities proportional to $q_0$, which are expressed relative to the amplitude prefactor
\be
  \zeta_0 = q_0\omega/g.
\ee 
With these,
\begin{subequations}\label{eq:znondim}
\begin{align}
  \tzeta =& \frac{\zeta}{\zeta_0}=\frac1i \lim_{\epsilon\to 0^+}\int_{-\infty}^\infty \frac{\rmd\kappa}{2\upi}e^{\rmi \kappa \tx-\rmi T}\frac{(\kappa-\kappa_c)\cosh\kappa(\tH-1)+\sinh\kappa(\tH-1)}{[\tG(\kappa) - \rmi  \epsilon\tilde\Phi(\kappa)](\kappa-\kappa_c-\rmi \epsilon\kappa_c)\sinh\kappa\tH},\\
  \tG =& \kappa-F^2\coth\kappa\tH - F^2\sigma; \\
  \tilde\Phi=&2F^2\coth\kappa\tH+F^2\sigma; \\
  \kappa_c =& -1/\sigma,
\end{align}
\end{subequations}
where $\kappa = kD$. 
We note the apprearance of the Froude number
\be
  F = \omega\sqrt{D/g}
\ee
based on velocity $\omega D$ and length $D$. 
Numerous examples of surface elevations are found in Fig.~\ref{fig:panels} (amplitudes are exaggerated beyond the linear regime in some cases, for visibility reasons).
Amplitudes $\tzeta$ from the full Euler theory derived here are compared to the corresponding results for potential flow \citep{tyvand14} in Fig.~\ref{fig:deepz}  for the upstream regular wave in deep water. The difference is moderate, and more pronounced for small $F^2$ and large $\sigma$. The corresponding quantity for downstream regular waves is plotted in Fig.~\ref{fig:downz}; clearly the downstream regular wave has radically different amplitude than for the potential solution, because much of the mass flux goes into the critical wave which is absent in the latter. Note that with the present theory, $\tzeta$ for downstream regular waves is a continuous function of $\sigma$ at $\sigma =1$ as would be expected physically. The amplitude $\tzeta$ for the critical wave is plotted in Fig.~\ref{fig:deepcrit}. Unlike the downstream regular wave, the critical wave remains nonzero for $\sigma>1$; its amplitude is always $\tzeta=1$ at $\sigma=1$. 

\begin{table}
  \begin{center}
  \begin{tabular}{lll}
  Dimensional quantity& Non-dimensional quantity& Name \\
   $S$ & $\sigma=S/\omega$ & Shear parameter \\
   $H$ & $\tH=H/D$ & Relative depth\\
  $\zeta$ & $\tzeta=\zeta/\zeta_0$ & Nondimensional elevation \\
   $x,z$ & $(\tx,\tz)=(x,z)/D$ &Nondimensional positions\\
   $t$ & $T=t\omega$ &Phase\\
   $k$ & $\kappa=kD$ & Wave number 
  \end{tabular}
  \caption{Table of non-dimensional quantities.}
  \label{tab:units}
  \end{center}
\end{table}

The far-field wave expressions now become, assuming $\kappa_-\neq\kappa_c$, 
\bs
\begin{align}
  \tzeta^>_\text{f.f.} =& \frac{ e^{\rmi \kappa_+\tx-\rmi T}}{\tG'(\kappa_+)\sinh \kappa_+h}\left[\cosh\kappa_+(\tH-1)+\frac{\sinh\kappa_+(\tH-1)}{\kappa_+-\kappa_c}\right],\\
  \tzeta^<_\text{f.f.} =& -\frac{ e^{\rmi \kappa_-\tx-\rmi T}}{\tG'(\kappa_-)\sinh \kappa_-\tH} \left[\cosh \kappa_-(\tH-1) + \frac{\sinh \kappa_-(\tH-1)}{\kappa_--\kappa_c} \right] \notag\\&
  - \frac{ e^{\rmi \kappa_c \tx - \rmi T}}{\tG(\kappa_c)}\frac{\sinh \kappa_c(\tH-1)}{\sinh \kappa_c \tH},\label{eq:zfDSnondim}
\end{align}
\es
and when $\kappa_-=\kappa_c$ we have
\begin{align}
   \tzeta^<_\text{f.f.} =-\frac{ e^{\rmi \kappa_c \tx-\rmi T}}{\tG'(\kappa_c)\sinh\kappa_ch} \left\{h\frac{\sinh\kappa_c}{\sinh\kappa_c h}+\left[\rmi \tx-\frac{\tG''(\kappa_c)}{2\tG'(\kappa_c)}\right] \sinh\kappa_c(h-1)\right\}
\end{align}
with
\begin{align}
  \tG'(\kappa_\pm) &= 1+\frac{\tH}{F^2}[(\sigma F^2-\kappa_\pm)^2-F^4], \\
  \tG''(\kappa_-) &= -\frac{2\tH^2}{F^{4}} (\sigma F^2-\kappa_-)[(\sigma F^2-\kappa_-)^2-F^4]. 
\end{align}

\begin{figure}
  \begin{center}
    \includegraphics[width = .9\textwidth]{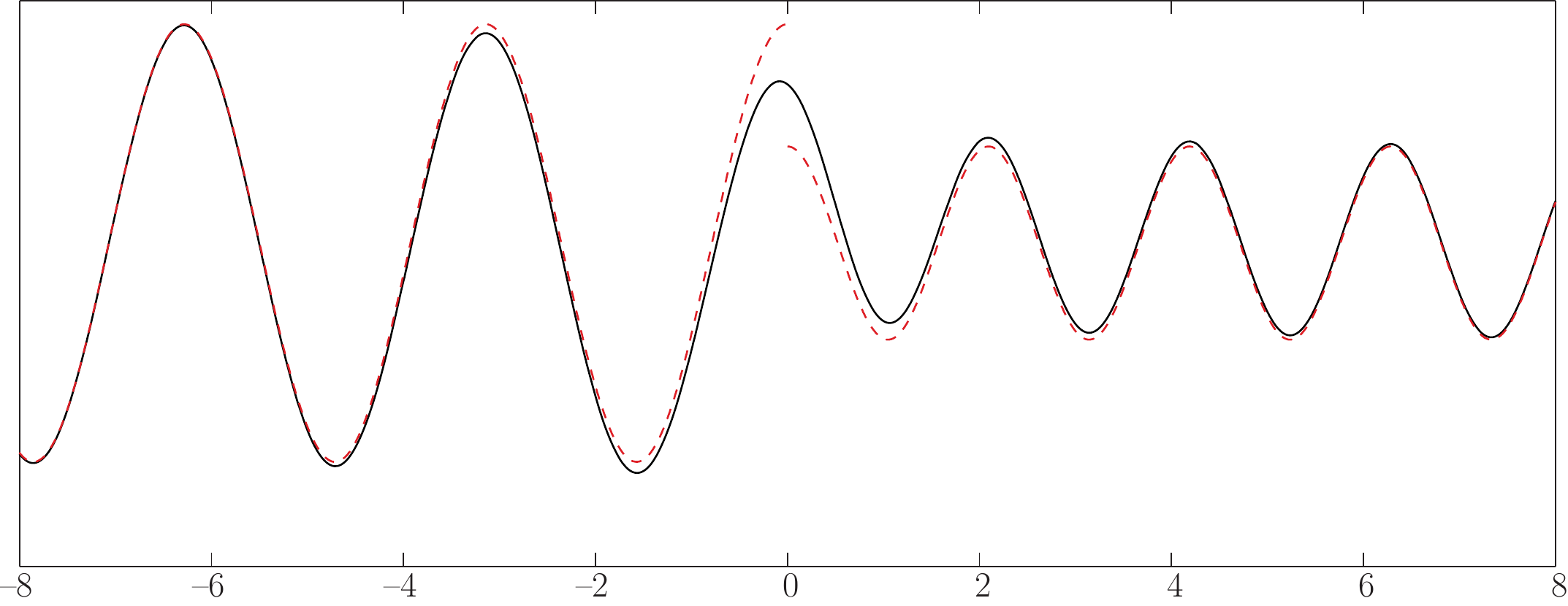} 
  \end{center}
  \caption{Example comparison between far-field solution (dashed line) and full solution (solid black line), here for deep water, $\sigma=0.5$ and $F^2=2$.  }
  \label{fig:farfull}
\end{figure}

In particular, the deep water asymptotic expressions in the far field are
\bs\label{ffdeep}
\begin{align}
  \tzeta^>_\text{f.f.} =&  e^{\rmi \kappa_+^\infty\tx-\rmi T}e^{-\kappa_+^\infty}\left(1+\frac1{\kappa_+^\infty-\kappa_c}\right)\\
  \tzeta^<_\text{f.f.} =&  e^{\rmi \kappa_-^\infty\tx-\rmi T}e^{\kappa_-^\infty}\left(1-\frac1{\kappa_-^\infty-\kappa_c}\right)\Theta(1-\sigma)
  +\frac{ e^{\kappa_c }e^{\rmi \kappa_c \tx -\rmi T}}{\kappa_-^\infty-\kappa_c}
\end{align}
\es
where
\be
  \kappa_\pm^\infty = F^2(\pm1+\sigma).
\ee

Figure \ref{fig:farfull} shows a comparison of the full solution, evaluated by numerical integration of \eqref{eq:znondim}, and the far-field asymptotics. As is clear, the far-field solution is a reasonable approximation as little as half a wavelength either side of the source.
In comparison the result obtained by treating the problem using a standard potential theory solution is \citep{tyvand14}
\begin{align}
  \tzeta^>_\text{f.f.,pot} =&  e^{\rmi \kappa_+^\infty\tx-\rmi T}e^{-\kappa_+^\infty}\\
  \tzeta^<_\text{f.f.,pot} =&  e^{\rmi \kappa_-^\infty\tx-\rmi T}e^{\kappa_-^\infty}\Theta(1-\sigma),
\end{align}
i.e., lacking the terms containing $\kappa_c$. In both upstream and direction we see that the amplitude of the propagating wave is modified, while in the downstream direction there is an additional critical wave whose phase velocity equals the current velocity at depth $D$.

\begin{figure}
  \begin{center}
    \includegraphics[width = .9\textwidth]{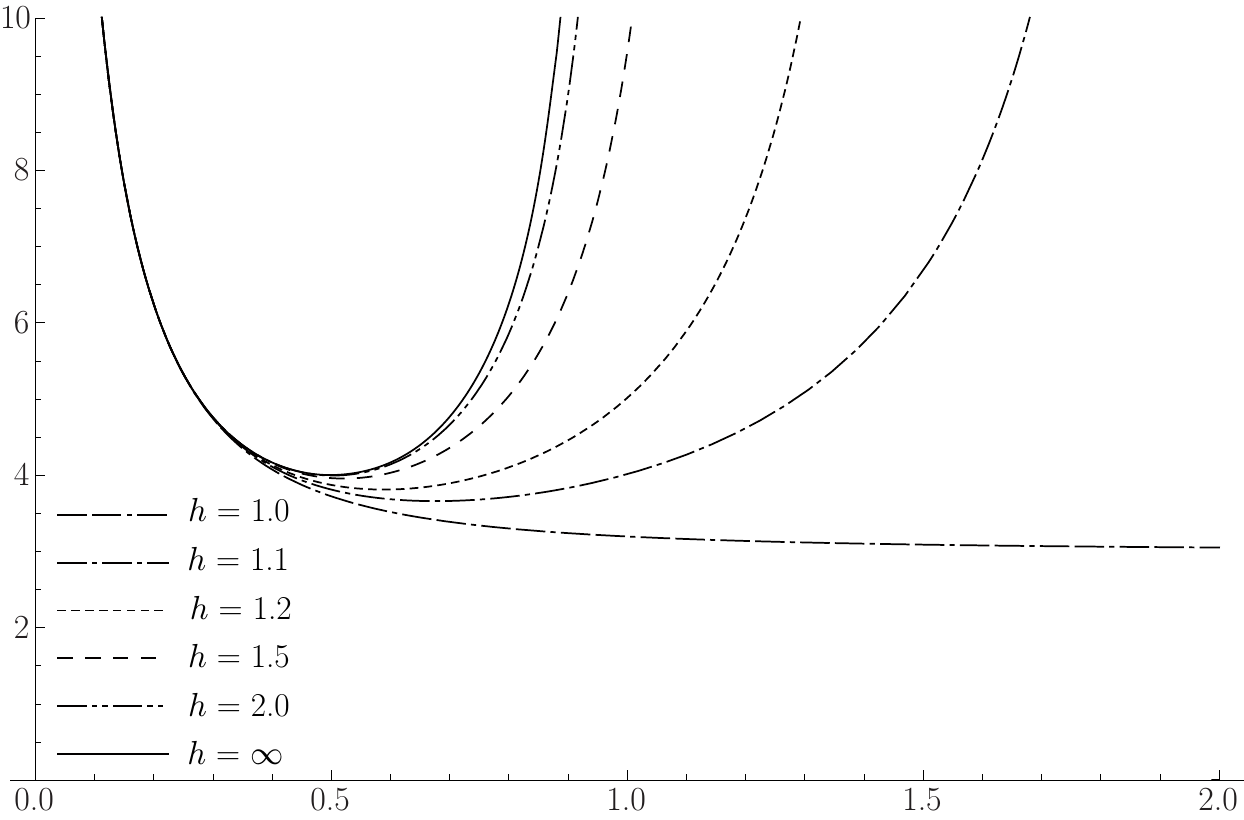} 
  \end{center}
  \caption{Resonant value of $F^2$ as function of $\sigma$ for different $\tH$.}
  \label{fig:resonant}
\end{figure}

\subsection{Critical wave and resonance condition}

As discussed in previous sections, when $\kappa_-\to\kappa_c$ the amplitude of the regular downstream and critical waves each diverge, and the two partly cancel each other. In deep water this resonant situation (which is quite different from the Doppler resonance studied, e.g., by \citet{tyvand15}) occurs when
\be
  F^2_\text{res}=\frac{1}{\sigma(1-\sigma)} ~~ \text{ or } ~~ \sigma_\text{res} = \frac12\pm\sqrt{\frac14-\frac1{F^2}}.
\ee
The situation can thus only occur when $F^2\geq 4$. In the case where $\kappa_-$ is close to $\kappa_c$ it is no longer meaningful to consider the two different waves separately since the overall downstream wave picture is dominated by the interference between the two. 
For finite water depth the resonant Froude number is found from Eq.~\eqref{eq:special} as
\be
  F^2_\text{res} = \frac1{\sigma[\coth(h/\sigma)-\sigma]},
\ee
while no explicit expression for $\sigma_\text{res}$ can be found in this case. The behaviour of the renonant value of $F^2$ for different $\sigma$ and $h$ is shown in Fig.~\ref{fig:resonant}. In particular, the minimum resonant value that can take $F^2$ is $3$, which is obtained in the limit $h\to 1$ and $\sigma\to\infty$.

\begin{figure}
  \begin{center}
    \includegraphics[width = \textwidth]{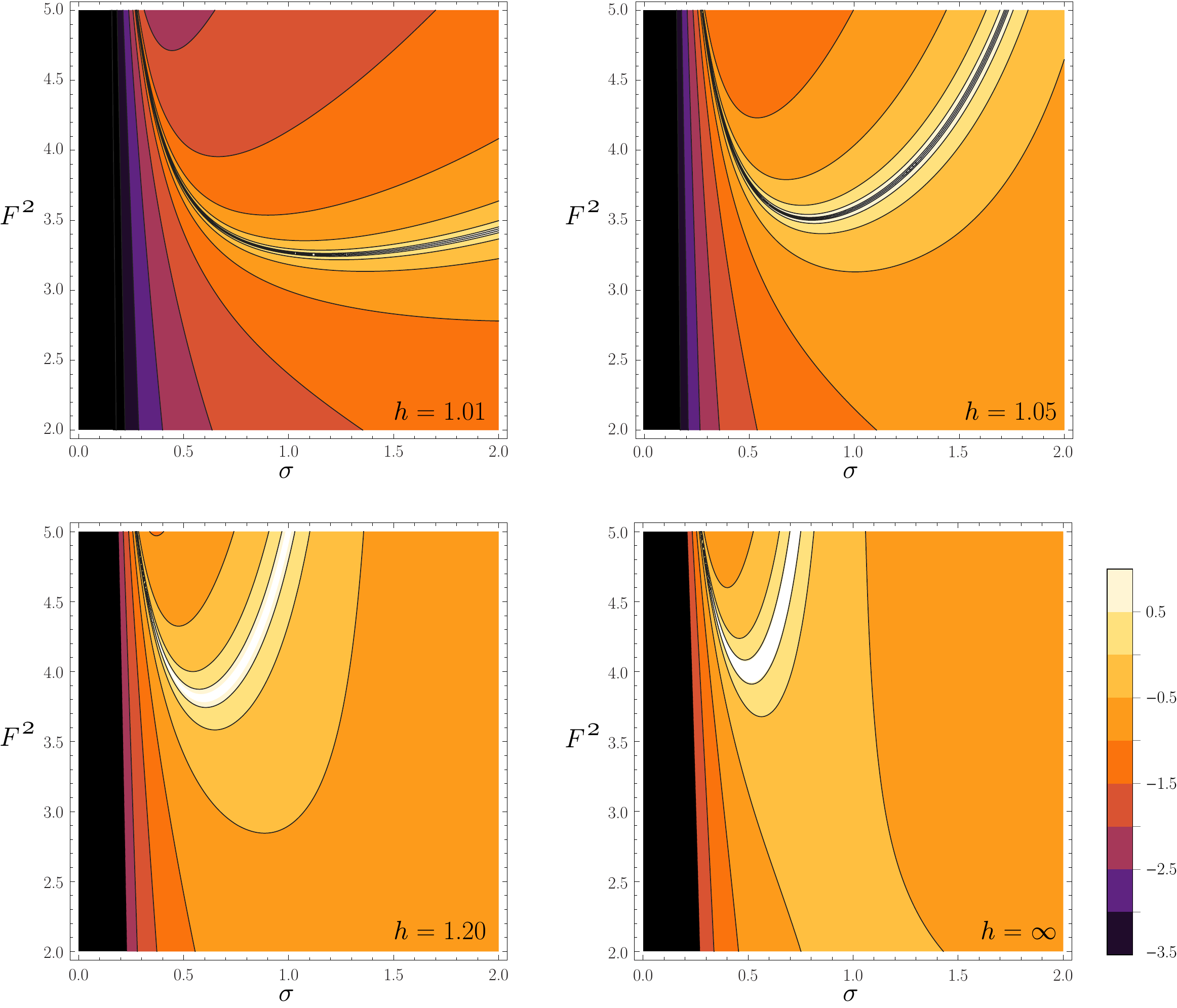} 
  \end{center}
  \caption{Contour plots of $\log_{10}|A_c|$ as given in Eq.~\eqref{eq:Ac}.}
  \label{fig:contourpanels}
\end{figure}

Let us turn to the amplitude of the critical wave, which in the far field can be found from Eq.~\eqref{eq:zfDSnondim} as
\be\label{eq:Ac}
  A_c =   
  -
  \frac{\sinh[(\tH-1)/\sigma]}{F^2\cosh(\tH/\sigma)-(1/\sigma+\sigma F^2)\sinh(\tH/\sigma)}.
\ee
We plot the logarithm of $|A_c|$ in Fig.~\ref{fig:contourpanels}. The critical curves shown in Fig.~\ref{fig:resonant} are clear to see as a ridge at which the amplitude diverges and changes sign. $A_c$ falls off extremely quickly as $\sigma\to 0$.
The behaviour for large $\sigma$ is particularly noteworthy. In this limit the amplitude $A_c$ is independent of $\tH$ except when the source is very close to the bottom ($\tH$ very close to $1$). Explicitly,
\be
  |A_c |
  \sim \frac1{F^2\sigma} + \frac{F^2+6\tH-3F^2\tH}{6F^4(\tH-1)}\frac1{\sigma^3} + \mathcal{O}(\sigma^{-5}).
\ee

\begin{figure}
  \begin{center}
    \includegraphics[width = \textwidth]{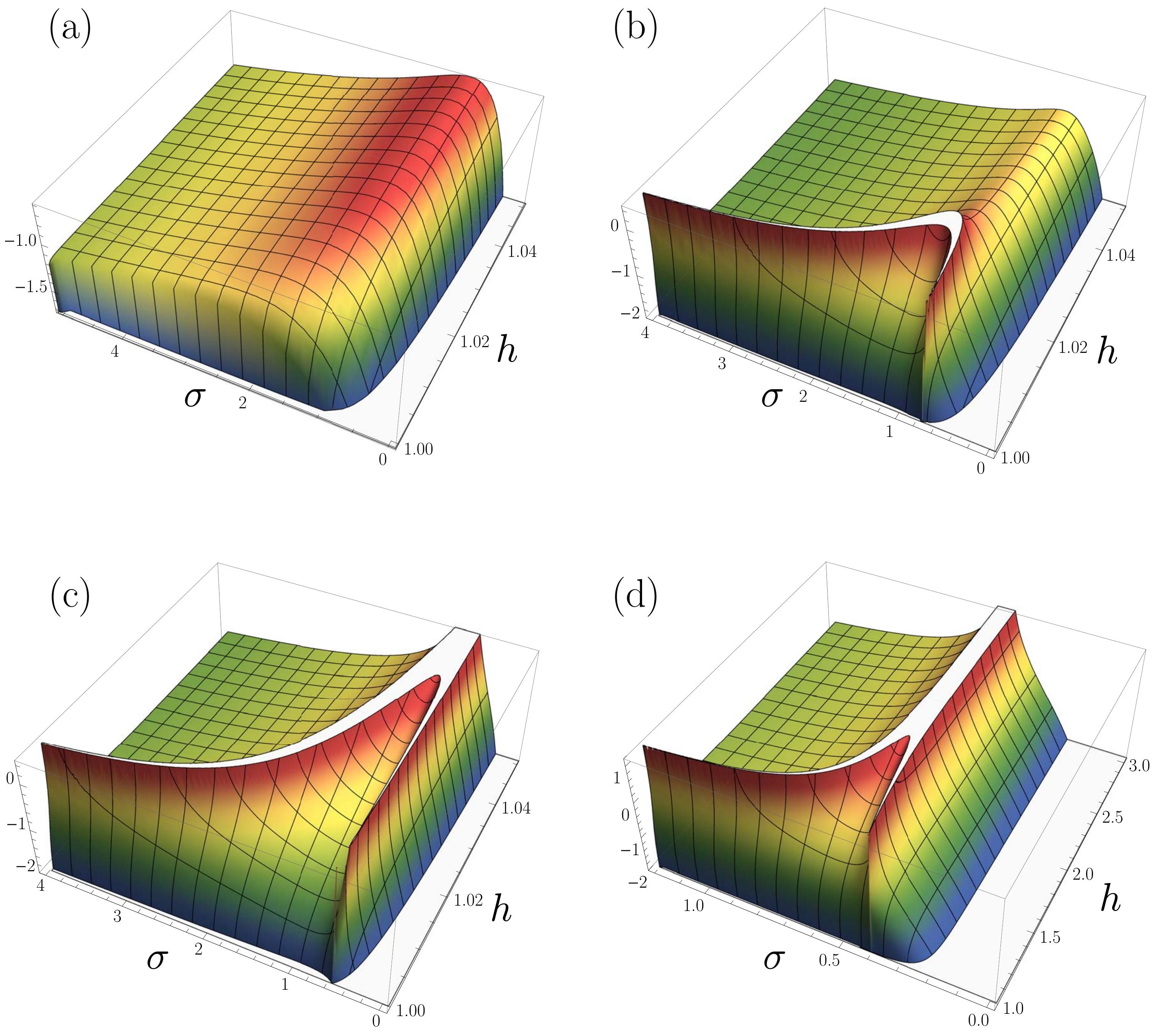} 
  \end{center}
  \caption{Plots of $\log_{10}|A_c|$ as given in Eq.~\eqref{eq:Ac} for $F^2 = 3$ (a), $3.2$ (b), $3.5$ (c) and $4$ (d).}
  \label{fig:3Dpanels}
\end{figure}

Finally we regard the behaviour of the amplitude of the critical wave for different values of $h$ and $\sigma$ when $F^2$ is constant. This is shown in Fig.~\ref{fig:3Dpanels}. In figure \ref{fig:3Dpanels}a, $F^2=3$ and so no resonance between the critical and downstream regular wave can occur. $A_c$ now behaves smoothly as $h$ and $\sigma$ are varied. Once $F^2$ exceeds $3$, resonance is possible, first for $h$ close to $1$ and, when $F$ reaches $4$, also in deeper waters. Again we notice that the behaviour is essentially as for infinitely deep water whenever $h>2$.

\subsection{Critical wave versus mass flux}

It is interesting to see whether the critical wave carries away mass flux that it receives from the source. The volume per length (area) emitted by the source during half a period is  $2q_0/\omega$. Similarly, the area below a wave crest with amplitude $Z_0$ and wavenumber $k$ is  $2Z_0/|k|$. We want to investigate how great a portion of the wave crest area emitted from the source during half an oscillation period actually goes into the crests of the far-field regular waves and the far-field critical wave. Let us call these fractions $f_0$ and $f_c$, for the regular waves and the critical waves, respectively, whereby
\bs
\begin{align}
  f_0 =& (\zeta_+/k_+ + \zeta_-/|k_-|)\omega/ q_0 \\
  f_c =& (\zeta_c/|k_c|) \omega/q_0
\end{align}
\es

In the hydrostatic shallow-water limit, it is known that $f_0 =1$ when there is no shear flow. Dispersion changes this ratio and makes it depend on the depth ratio $H/D$. In the present context we will not discuss the variation of $f_0$, but focus on the behavior of $f_c$, since the critical wave can be suspected for removing a considerable amount of the mass flux that should otherwise have gone from the source into the regular waves. The general formula for $f_c$ is 
$f_c = \sigma F^2 A_c$, 
and it may be plotted in a similar way as the dimensionless elevation of the critical wave. $f_c$ will be greater than one in a parameter domain near resonance, since $f_c$ tends to infinity at resonance, and will be of order one or greater unless either $F^2 < 2.5$ or $\sigma$ is considerably smaller than one. When $\sigma$ is large, $f_c$ tends to settle at values close to one, with relatively weak dependence on $F^2$. This happens in a parameter domain quite far from resonance, which implies that large portion of the source flux for strong shear rates will go directly into the critical wave, to be convected away with the flow. In this parameter domain it can be expected that the critical layer will reduce the amplitudes of the regular waves considerably.

In potential flow from oscillatory sources, the input of potential energy from the source to the traveling waves is an important mechanism for outward energy flux. The critical layer-like flow steals some of the source mass flux that would otherwise have gone into the regular wave elevations.  This stolen mass flux is visible as crests and troughs convected with the flow velocity at the source depth. The process approaching resonance is different because the critical wave and the regular wave will compose a joint interference wave that has an envelope growing linearly in space near its node points. Near resonance it is no longer meaningful to compare the convected mass flux of the critical wave with the emitted mass flux from the source.

\section{Flow pattern: downstream vortical structures}\label{sec:velocity}

We finally consider the sub-surface velocity field, with special emphasis on the vorticity generated by the oscillating source and associated downstream vortical structures. 

\begin{figure}
    \includegraphics[width = \textwidth]{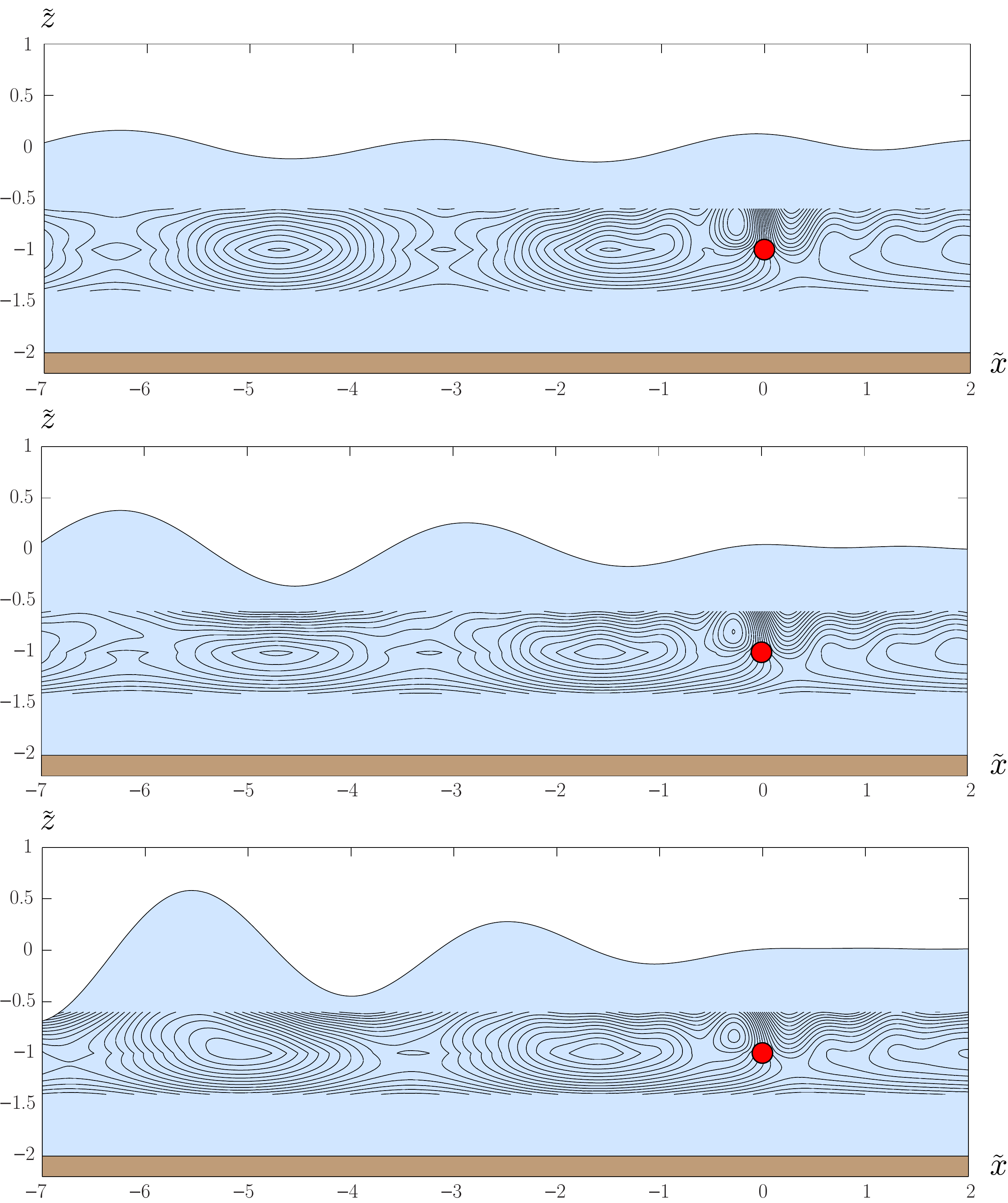}
  \caption{Example wave patterns, and streamlines as measured from a system moving downstream (towards the left) at velocity $SD$ for $F^2=2$ (top), $F^2=3$ (middle) and $F^2=4$ (bottom). Here $\zeta_0/D=1,\sigma=0.5$ and $H=2$. The source is shown as a filled circle, and the snapshot is taken at $t=0$. The bottom plot shows the situation close to resonance where downstream wave amplitude diverges linearly away from the source. (See also video in supplementary material)}
  \label{fig:streamlines}
\end{figure}

From the calculations in Section \ref{sec:solution} we can find the two velocity components in the spectral domain
\begin{align}\label{eq:vels}
  \ba{-\rmi u}{w} =& A\ba{\cosh}{\sinh} \kappa(\tz+\tH) \notag \\
  &+ \left[\ba\sinh\cosh \kappa(\tz+1)-\frac1{\kappa-\kappa_c}\ba\cosh\sinh \kappa(\tz+1)\right]\Theta(\tz+1),
\end{align}
where
\begin{align}
  A=&\frac1{\tG(\kappa) }\left[\left(F^2+\frac{\kappa-\sigma F^2}{\kappa-\kappa_c}\right)\frac{\sinh\kappa}{\sinh\kappa\tH} \right. 
 -\left.\left(\kappa-\sigma F^2+\frac{F^2}{\kappa-\kappa_c}\right)\frac{\cosh\kappa}{\sinh\kappa\tH} \right].\label{A}
\end{align}
This expression is not particularly useful numerically when $\tz>-1$, however, since in this case each of the terms in the square brackets in \eqref{eq:vels} give diverging Fourier integrals. By using the boundary conditions in Eqs.~\eqref{eq:boga} and \eqref{eq:aogb} we may write velocities instead as
\begin{align}
  \ba{-\rmi u}{w} =& \rmi \omega B \left[
  \left(\sigma-\frac\kappa{F^2}\right)
  \ba{\cosh}{\sinh} \kappa \tz-\ba\sinh\cosh \kappa\tz\right],
\end{align}
valid when $\tz>-1$, with 
\be
  \rmi \omega B = \frac{F^2}{\tG\sinh\kappa \tH}\left[\cosh\kappa(h-1)+\frac{\sin\kappa(h-1)}{\kappa-\kappa_c}\right]
\ee
as found previously.

We can also write down the stream function for the full flow (in nondimensional form) which is zero at the bottom and continuous at $\tz=-1$,
\be
  \tilde\Psi(\tx,\tz) =\half\sigma(\tz^2-\tH^2)+\frac{\zeta_0/D}{F^2}\int_{-\infty}^\infty\frac{\rmd\kappa}{2\upi}\psi(\kappa,\tz)e^{\rmi \kappa\tx-\rmi T}
\ee
where $\tilde\Psi = (\omega D^2)^{-1}\Psi$ and 
\begin{align}
  \psi(\kappa,\tz) =& \frac{\rmi A}{\kappa}\sinh\kappa(\tz+h)+\frac \rmi \kappa 
  \Bigl[\cosh\kappa(\tz+1)
  \notag \\
  &-\frac{\sinh\kappa(\tz+1)}{\kappa-\kappa_c}\Bigr]\Theta(\tz+1).
\end{align}
Again, a more useful form for $z>-1$ is
\be
  \psi(\kappa,\tz) = \frac{\rmi }{\kappa}
  \left[\left(\sigma-\frac\kappa{F^2}\right)\sinh\kappa\tz-\cosh\kappa\tz\right]
   \rmi \omega B.
\ee
To evaluate the integral, the poles at $\kappa_\pm$ and $\kappa_c$ are moved slightly off the real axis by means of the radiation condition exactly as in previous sections.

An interesting pattern emerges when we regard streamlines as seen from a system moving downstream at velocity $SD$, i.e., the shear flow velocity at the depth of the source. In this system we find the characteristic closed contours associated with the cat's eye vortices of a critical layer, of the kind previously found by \citet{ehrnstrom08} and \citet{wahlen09}. 

In order to plot streamlines in the system which moves dowstream with the speed at which the 
critical layer-like flow structures are
advected, we add the constant motion to the velocity profile
\be
  \tilde\Psi_\text{advected} = \tilde\Psi + \sigma\tz.
\ee
The resulting streamline patterns near the critical layer (at $\tz=-1$) are shown in Fig.~\ref{fig:streamlines} together with the corresponding surface elevation for increasing values of the Froude number $F^2$. The closed streamlines are clear to see, forming structures reminiscent of the cat's eye vortices typically associated with the Kelvin-Helmholtz instability, which is the underlying process also here. A video rendition of these plots shows how the vortices are advected downstream, their position remaining underneath or slightly ahead of a corresponding wave trough.

For the parameters in the Fig.~\ref{fig:streamlines} ($\tH=2, \sigma=0.5$) the resonant value is $F^2_\text{res}=3.99$, so the bottom panel essentially shows this situation. When the Froude number is a little higher or a little lower than $F^2_\text{res}$, the downstream wave forms the well known slowly modulated pattern when plane waves of almost equal wavelength and amplitude are superposed. 

The vorticity $\Omega$ in the flow can be calculated from the perturbation velocities and only obtains a nonzero contribution from the derivative of the Heaviside function at $\tz=-1$. We find, after solving the resulting Fourier integral by using the radiation condition the same way as before using the contours in Fig.~\ref{fig:contours}
\be\label{eq:omega}
  \Omega = \nabla\times\bi{v} = S+\frac{\partial \hat{u}}{\partial z}-\frac{\partial \hat{w}}{\partial x} = S - \frac{q_0}{D^2} \rme^{\rmi \kappa_c \tx-\rmi T}\delta(\tz+1)\Theta(-\tx).
\ee
where the vectorial vorticity is $\Omega\bi{e}_y$. Hence the vorticity vector has the same wave form as the critical wave itself and the vorticity field is concentrated in a thin sheet at the source depth which is convected downstream along with the vortices indicated by the streamline graphs in Fig.~\ref{fig:streamlines}. 
While the vorticity \eqref{eq:omega} comes out of a lengthy Fourier-transform calculation of the velocity field, it might have been derived simpler directly from vorticity equation \eqref{eq:vort}. It is straightforward to verify by insertion that \eqref{eq:omega} satisfies the vorticity equation \eqref{eq:vort}. We show in Appendix \ref{app:crit} how the same vorticity results from a much more elementary consideration, and may be seen as a direct consequence of Kelvin's circulation theorem. 

A striking property of the additional perturbation term in \eqref{eq:omega} proportional to $\delta(\tz+1)$, is that it depends weakly on $S$ through $\kappa_c$ only, and does not vanish in the limit $S\to 0$. The latter point is particularly surprising, since the source does not in itself produce any vorticity, but can only modify the vorticity already present in the background flow in a manner which is made clearer in Appendix \ref{app:crit}. The reason for this peculiarity may be seen from considering the vorticity equation \eqref{eq:vort}. When a tiny material loop centered at depth $-D$ will spend a very short time proportional to $1/U(-D)$ with the source inside of it, hence will pick up additional vorticity proportional to $q_0 S\exp(-\rmi \omega t)/U(-D)$. However, since $U(-D) = -SD$, this prefactor becomes independent of $S$. Clearly this is a very special case which only occurs for a perfect Couette profile $U(z)=Sz$ when the source is at rest relative to the surface. Should even a small relative velocity between source and surface be present, or the shear profile be anything other than a perfect Couette, the perturbation vorticity will vanish as $S\to 0$.

\subsection{Remarks on the critical layer-like flow}\label{sec:critlayer}

The flow picture just discussed is reminiscent of a critical layer, known to exist for surface waves in the presence of shear flow when the phase velocity of a wave equals the sub-surface flow velocity at some critical level \citep[e.g.][]{drazin81}. Our flow has a particular, ```critical'' wave solution, whose velocity equals the flow velocity at the source depth $D$, which is also the advected velocity of the vortical flow structures. For this reason we have referred to this wave as the critical wave, and its associated sub-surface flow pattern as a critical layer. 

There exists a large literature on the various important manifestations of critical layers (see also Introduction), and in this context it is important to point out an important difference between ``standard'' critical layers, such as they appear, e.g., in the study of Rossby waves \cite[e.g.][]{killworth85}, and that present here. In the example of critical layers in Rossby waves, the critical layer comes into existence when a wave generated by a forcing elsewhere, sets fluid in motion at the critical level where flow velocity exactly equals the wave's phase velocity. The critical level thus follows from the dispersion relation and the shear profile. 

In the present case, however, the critical layer-like vortex flow is generated at the level where the source is situated, and is caused by the source itself. The spatial periodicity of the vortical structures is given by the source frequency and the local flow velocity near the source, independently of the dispersion relation for the regular gravitational waves. The critical layer is convected at the local flow velocity at depth $D$. This speed of a critical wave is not equal to the phase velocity of propagating waves except for a single ``resonant`` forcing frequency, at which a resonance occurs, as analysed above. The free surface is necessary for the vortex structure to form, but once created, these alternating vortices drift downstream with the flow, and the critical surface wave drifts with it as the surface manifestation coupled with these sub-surface structures. After their generation, the critical layer and associated critical wave do not interact further with the regular dispersive wave which is also being forced by the same source.

Our purpose herein has been to lay out the properties such as they are of a basic building block for flow modelling, the strictly linearized (i.e., infinitesimal amplitude) oscillating source in the presence of a shear current, from which more complex flow pictures might be constructed via superposition. One might note that the solution does not necessary represent well the flow from a realisable, \emph{finite} amplitude source, whose generated vortices could well interact to form flow patterns different from linear theory predictions. Indications to this effect are found for somewhat similar flows in Rossby wave critical layers, where linear solutions \citep{dickinson70,warn76} tend to develop non-linearly in time \citep{stewartson77,warn78}, and are unstable \citep{killworth85}. It may be considered a limiting case of a localised vorticity disturbance, whose stability was studied by \citet{balmforth97}. The assumptions of linear theory are clearly broken at resonance, as we have shown above. We do not pursue such detailed considerations here, however, firstly because a strictly linear theory is necessary for superposition and eventual deployment as modelling tool, and secondly because the modelling of bodies in flow will most likely require cancellation of the critical layer-type far-field solution in any case, as discussed in the next Section.

\subsection{On the possibility of critical layer-free singularities}\label{sec:critfree}

As discussed in the Introduction, a main motivation for studying oscillating sources in the presence of a shear current is the hope to eventually develop a theory to allow the study of waves interacting with floating or submerged bodies, similar to the theory for irrotational flow which has been so successful in marine hydrodynamics \cite[c.f., e.g., ][]{faltinsen90, newman77}.
The discovery that an oscillating source produces a 
critical layer-like downstream flow pattern, 
including an undulating vorticity sheet in its wake, is bad news in this regard, at least at the outset. 

In potential theory, Rankine-type bodies are made by simply inserting (time dependent) sources and sinks into a flow, yet naively interpreted, such a procedure would create a vorticity sheet in the wake of the modelled ``body'' when vorticity is present. In 2D inviscid flow the vorticity of each fluid particle is conserved as it moves past a submerged body, however, so no such downstream vorticity sheet can be allowed. 

We do not provide a resolution of this complication at present; The general question of whether and how a moving body in a shear flow may be satisfactorily described by a sum of flow singularities is an open question to
be investigated carefully in future works. We do note in passing, however, that it is possible for a downstream source to cancel the critical layer from an upstream one, 
and that such a ``critical-layer free" pair of sources can if desired be merged into a source-doublet combination in a single point. 

To wit, using two oscillating sources aligned streamwise and with the appropriate phase difference, we can easily construct a modified dipole which does not create a critical layer. Any perturbation quantity $A(x,t)$ (e.g., velocity, stream function, surface elevation) surface  in our theory is proportional to $q_0$ and has the general form (a possible $z$ dependence is understood)
\be
  A = q_0 \int_{-\infty}^\infty \frac{\rmd k}{2\upi} a(k) \rme^{\rmi k x - \rmi T}.
\ee
Now adding a second source a little downstream, at $x=-\Delta x$, one readily ascertains that the additional vorticity \eqref{eq:omega} vanishes at positions $x<-\Delta x$ if the second source has an appropriate phase shift $\alpha = k_c\Delta x + \upi$ (note difference from a standard doublet, whose phase difference is $\upi$):
\begin{align}\label{NCL}
  A_\mathrm{clfd} =& q_0 \int_{-\infty}^\infty \frac{\rmd k}{2\upi} a(k) [\rme^{\rmi k x - \rmi T} + \rme^{\rmi k (x+\Delta x) - \rmi (T+\alpha)}]\notag \\
  \buildrel{\Delta x\to 0}\over{\longrightarrow}& \rmi D_0 \int_{-\infty}^\infty \frac{\rmd k}{2\upi} a(k) (k-k_c) \rme^{\rmi k x - \rmi T}
\end{align}
where $D_0 = q_0 \Delta x$ is the strength of the resulting modified doublet. The prefactor $k-k_c$ now eliminates the pole at $k=k_c$, and there is no critical layer contribution to the far-field. Eq.~\eqref{NCL} describes a combination of a point source of strength $-\rmi k_cD_0$ and a doublet directed along the $x$-axis.

For example, the deep water far-field waves from such a modified doublet are
\bs
\begin{align}
  \tzeta_\ff^> =& \rmi\Delta x\, \rme^{\rmi \kappa_+^\infty\tx-\rmi T}\rme^{-\kappa_+^\infty}(\kappa_+^\infty - \kappa_c + 1) \\
  \tzeta_\ff^< =& \rmi \Delta x\, \rme^{\rmi \kappa_-^\infty\tx-\rmi T}\rme^{\kappa_-^\infty}(\kappa_-^\infty - \kappa_c - 1)\Theta(1-\sigma)
\end{align}
\es
in generalisation of Eq.~\eqref{ffdeep}. Note that these waves still have different amplitudes from those predicted by \cite{tyvand14} for the same source/doublet combination.


\section{Concluding remarks}

The water wave radiation problem for a submerged oscillating line source has been solved analytically in the presence of a shear flow. The fluid depth $H$ is constant, the oscillating source is at rest with respect to the undisturbed surface and the chosen coordinate system. The mass flux from the source generates a critical layer-like vortical flow structure downstream of the source at the source's depth, $D$, which we refer to simply as the critical layer. This layer generates a third ``critical'' wave in addition to affecting the amplitudes of the regular upstream and downstream waves, respectively. The critical wave is non-dispersive, and its velocity is the shear flow velocity at the submergence depth of the source; it is the surface manifestation of an underwater phenomenon. Indeed, when seen from a system moving downstream at the fluid velocity at the source depth, the streamlines form closed loops which are advected downstream, in a manner akin to Lord Kelvin's cat's eye vortices. 

The critical layer is an underwater phenomenon being modified by gravity at the free surface, while the regular upstream and downstream waves are gravitational surface waves modified by the critical layer. When the wavelength of the critical wave approaches that of the downstream regular wave, a resonance is found to occur between the two, resulting in a combined downstream wave whose amplitude increases indefinitely as a linear function of the distance from the source. 

When the oscillating source is placed at the bottom instead of in the fluid domain, the critical layer disappears. 
Also, for a bottom source, streamlines are unable to form closed loops as described above. On the other hand, we have shown that the critical layer is very important even for $H/D$ only a few per cent above $1$, as long as the squared Froude number exceeds $3$. Unless the Froude number is small and the shear is weak, there is no smooth transition from the full hydrodynamic model described by the Euler equation to the potential flow model valid for a bottom source.  

At the source position, the source flow and the rotational background current combine to produce a thin sheet of vorticity which is swept downstream and whose sign and magnitude oscillates in the same manner as the critical wave itself. This is not in contradiction to Kelvin's circulation theorem for inviscid flow, which states that the circulation around a closed path passively advected must be constant, but rather a consequence of it; 
a downstream drifting curve which encircles the source for a brief time will have its encircled area modified as the source injects or extracts fluid from its interior, and since its circulation stays the same, its vorticity must change. 
The result to be expected 
is a series of counter-rotating fluid structures being advected with the flow, which is exactly what we observe when we study the streamlines of a velocity field as seen from an intertial frame moving downstream at velocity $SD$, the fluid velocity at the source's submergence depth. 
This simple argument is formalised in Appendix \ref{app:crit} where it is showed how such simple considerations are able to reproduce the exact vorticity distribution found in our full-fledged analysis. Remarkably, the amplitude of the perturbation vorticity is independent of the basic flow vorticity. 
Much the same might have been concluded based on simple arguments by directly regarding the 2D vorticity equation, which shows that vorticity is conserved for each particle \emph{except} for particles passing through the source point.

The 
critical layer-like vortical flow
is a necessary effect caused by any time-dependent submerged source submerged in a shear flow with nonzero throughflow through the singular source point. With zero throughflow in the source point, there would be no critical layer, because the generated perturbation vorticity needs to be carried away from its source. The throughflow through a simple submerged source point therefore poses additional challenges for the use of such singularities as Green functions for describing rigid bodies that are floating or submerged in a shear flow. Our physical arguments in Appendix \ref{app:crit}, 
as well as consideration of the vorticity equation in Sec.~\ref{sec:introvort}, clearly demonstrate that the generated vorticity is a necessary aspect of submerged sources in shear flows, and that 
the potential flow method for the same problem employed by \cite{tyvand14} 
(which does not predict a critical layer) is physically inconsistent.

Clearly, in inviscid flow, the presence of a body cannot introduce the additional vorticity associated with the critical layer, hence, if a distribution of sources is to model such a body, it is necessary that said distribution must not only satisfy applicable boundary conditions on the body's surface but also either re-absorb its own generated downstream vorticity or be constructed so that generated vorticity is not convected into the outside flow. Solving such problems is a task for the future and no solution is provided at this stage. We note however, that it is possible to construct a tuned source/doublet combination which does not produce a downstream critical layer.

\subsection*{Acknowledgement}

The authors would like to thank Professor Christian Kharif for suggesting we revisit this problem with a fundamental approach.

\appendix

\section{Dispersion relation and cut-off in the deep water limit}\label{app:disprel}

Consider the dispersion relation \eqref{eq:disprelasjon} in the deep water limit:
\be\label{eq:appA}
     \varkappa = a \coth \varkappa+ b,
\ee
where we let $\varkappa = kH$ and define the shorthand $a= \omega^2 H/g$ and $b= \omega SH/g$. As $H\to\infty$, both $a$ and $b$ in general diverge. A wave of standard appearence will have $|k|$ of order $1/D$, and so $|\varkappa|$ of order $H/D\gg 1$, so we shall seek such a solution, for which we may assume $\coth\varkappa \approx \sg(\varkappa)$. Then
\be\label{eq:appB}
  \varkappa = \pm a + b.
\ee
As long as $a>b$ (i.e., $\omega>S$) we obtain one solution of either sign, which is a satisfactory solution. All terms of \eqref{eq:appB} are now proportional to $H$, and the resulting waves are depth independent as expected. If $a<b$ (i.e., $\omega<S$), however, both solutions to \eqref{eq:appB} are positive, and we must seek a different negative valued solution. Presume now the solution be $\varkappa \sim \mathcal{O}(1)$, in which case the terms containing $a$ and $b$ will dominate equation \eqref{eq:appA}, and we obtain instead $\coth\varkappa \approx -b/a=-S/\omega$, or
\be  
  k_- \approx \frac1H \arcoth\frac{S}{\omega} = \frac1{2H} \ln \frac{S+\omega}{S-\omega}
\ee
which is real and negative provided $S>\omega$, and tends to zero as $H\to \infty$.

Consider now the contribution from $k_-$ to the far-field expression of $\zeta$ in the downstream direction, from Eq.~\eqref{eq:zfdownstream}:
\be\label{eq:zetacutoff}
  \frac{g}{q_0\omega}\zeta_{\text{f.f.},-}\approx \frac{e^{-\rmi \omega t}}{\Gamma'(k_-)}\left(\coth k_-H -\frac S\omega \right)\approx \frac{2g}{H\omega S}e^{-\rmi \omega t}
\ee
where we used
\begin{align*}
  \Gamma'(k_-) =& 1+\frac{\omega^2H}{g\sinh^2k_-H}=1-\frac{\omega^2H}{g}(\coth^2k_-H -1) \\
  &\approx 1-\frac{\omega^2H}{g}(\frac{S^2}{\omega^2} -1) \approx -\frac{HS^2}{g}
\end{align*}
for $H\to\infty$, where $S>0$ is presumed. Eq.~\eqref{eq:zetacutoff} shows the amplitude of the $k_-$ wave vanishes as $1/H$ in the deep water limit. 

The energy transported by a simple monochromatic wave is proportional to the amplitude square and the group velocity. The dispersion relation \eqref{eq:disprelasjon} may be written with respect to $\omega$ as
\be
  \omega(k) = \sqrt{gk\tanh kH + (\half S\tanh kH)^2} - \half S\tanh kH.
\ee
When $k\to 0$ while $kH\sim \mathcal{O}(1)$, one has asymptotically
\be
  \omega(k) \sim \begin{cases} 2gk/S^2, & k>0, \\ -S\tanh kH, &k<0\end{cases}
\ee
and the group velocity therefore
\be
  c_g = \frac{\rmd \omega}{\rmd k} \sim \begin{cases} 2g/S^2, & k>0, \\ -\frac{SH}{\cosh^2 kH}, &k<0\end{cases}.
\ee
Thus, with amplitude proportional to $1/H$ and $c_g$ proportional to $H$ in the downstream direction, the energy transport becomes proportional to $1/H$ and vanishes in the limit of deep water. 

We conclude that the far-field wave from $k_-$ is simply cut off whenever $S>\omega$: Its wavelength is infinite, and we have ascertained that it carries no energy in said limit, hence may simply be ignored.

\section{Generation mechanism for the critical layer-like vorticity}\label{app:crit}

\begin{figure}
  \begin{center}\includegraphics[width=3.5in]{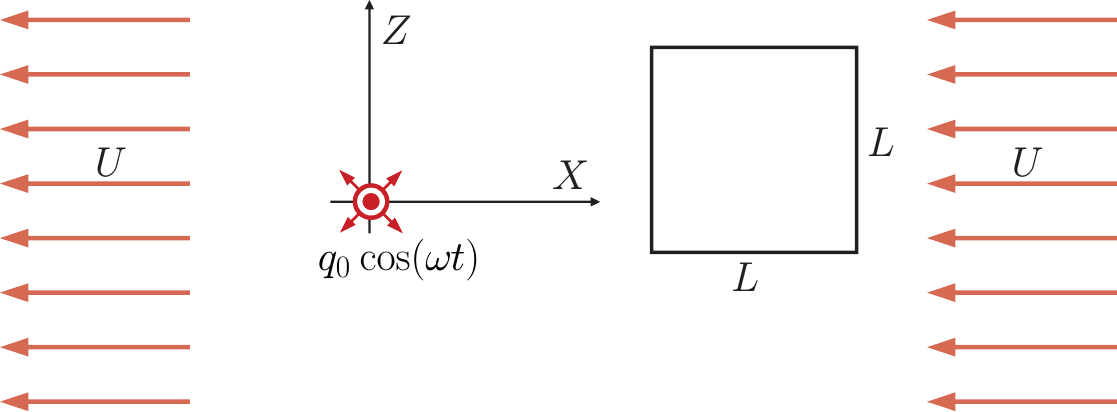}\end{center}
  \caption{Sketch of local configuration in an $X, Z$ plane. A small, initially rectangular material curve is advected with the flow, to be slightly deformed by the source.}
  \label{fig:curve}
\end{figure}

We will now study the physical mechanism of generating a downstream critical layer from an oscillating source submerged in a shear flow.
In order to do that, we will apply Lord Kelvin's circulation theorem to the evolution of vorticity for a small material curve that comes inside the influence of the source. This is a local analysis for the fluid domain around the source position, and we introduce local Cartesian coordinates $(X, Z)$ in two dimensions. The free surface and the bottom are both disregarded because they appear as distant and are unimportant for this argument. In this local analysis for the fluid domain around the source position, the basic flow velocity can be given a constant value $U$, essentially representing the shear flow velocity at the source position. The vorticity in the basic flow is $S$. See the sketch in Fig.~\ref{fig:curve}, 
where we have drawn a material curve that is going to be influenced by the line source. The 2D source is given a flux amplitude $q_0$ (area per time), and this flux is assumed infinitesimally 
small in comparison to the basic flow variables, to esure that linear theory is generally valid.

The material curve starts as exactly quadratic with side $L$, and it will be only slightly modified by the source flux $q(t)= q_0 \cos (\omega t)$. According to Lord Kelvin's circulation theorem, all material curves will have a circulation $\Gamma$ that is constant in time, and is equal to $S$ times the initial area inside the material curve, restricting this analysis to 2D material curves. The length of the rectangle must obey the restriction 
$L \ll |U|/\omega$, 
so that the time that a contour spends with the source located inside it is much smaller than an oscillation period. Then we need only to consider one instant $t=\tau$ as representative for the modification of the material curve due to the source. The instant $t=\tau$ is the midpoint of the interaction time interval during which the source is inside this particular material curve. This means that the area of the considered material curve is modified by a constant area flux $q_0 \cos (\omega \tau)$, and since the size of the interaction time interval is 
$L/|U|$, 
we find the following formula for the downstream value $A$ of the area inside the material curve
\be
  A = L^2 + \frac{q_0 L}{|U|} \cos (\omega \tau) \Theta(L/2 + Z) \Theta(L/2 - Z) \Theta(XU-L|U|/2),
\ee
where $(X,Z)$ now represents the midpoint of the material curve under consideration. The factor $\Theta(XU-L|U|/2)$ ensures that the volume is only altered if the material curve 
is on the downstream side and has finished its interaction with the source
(times during the very short period of interaction are not considered).
The upstream value of the area inside the material curve is $L^2$ by definition.

Lord Kelvin's circulation theorem now implies that
\be
  \Gamma = S L^2 = \Omega A,
\ee
where we have applied Stokes' theorem for the vorticity inside a small contour. 
Here $\Omega$ is the downstream vorticity inside the material curve, after having been modified by the source. Inserting the formula for $A$ then gives the approximate formula for the downstream vorticity inside a material curve
as $L \rightarrow 0$,
\be
  \Omega = S\left[1 - \frac{q_0}{|U|} \cos (\omega \tau)~ \delta(Z)  ~\Theta(U X) \right],
\ee
where we have introduced Dirac's delta function. 
The expression is valid to first order in the linearisation parameter $q_0/(|U|L)$.

Using the original shear flow velocity $U=-S D$ at the location of the source, the position of the square at a time $t>\tau$ is $X=-SD(t-\tau)$. 
If we now introduce the original 
coordinates
$(X, Z-D) = (x,z)$ 
we get the final formula for the vorticity distribution everywhere in the fluid
\be
  \Omega(x, z, t) = S - \frac{q_0}{D} \cos \left[\omega\left(-\frac{x}{SD} - t\right) \right]  \delta(z+D)  ~\Theta(-x),
\ee
which is equivalent to Eq.~\eqref{eq:omega}.

Obvious synchronization occurs regularly at discrete wavelength distances from $x=0$.
It is remarkable that the perturbation vorticity generated by the oscillating source has an amplitude that is independent of the shear rate $S$, since the existence of the perturbation vorticity requires that $S \ne 0$. 
In the entire fluid, except for the single singular point $(x,y) = (0, -D)$, the vorticity is conserved for each individual particle
according to Helmholtz theorem for vorticity evolution in 2D.
This explains the critical wave of oscillating vorticity travelling with the shear flow, but it gives no estimate for the amplitude of the critical surface wave. The wave amplitudes can be found by the full analysis above.

From our physical arguments a more general but qualitative result follows: Any time-dependent submerged source 
will modify the local vorticity in a shear flow, by changing the vorticity of the fluid particles that move through the source point while the source strength is nonzero.

\bibliographystyle{jfm}
\bibliography{wave}

\end{document}